\pdfoutput=1
\documentclass[a4paper,fleqn,usenatbib,usedcolumn]{mnras}

\usepackage{mathptmx}

\usepackage[T1]{fontenc}
\usepackage{ae,aecompl}

\usepackage{graphicx}	
\usepackage{amsmath}	
\usepackage{amssymb}	

\newcolumntype{\decol}{D{.}{.}{2}}

\title[Occurrence of BEPs in the classical instability strip]
{The occurrence of Binary Evolution Pulsators in the classical instability strip of RR~Lyrae and Cepheid variables.}

\author[P. Karczmarek et al.]{
P.~Karczmarek$^{1}$\thanks{E-mail: pkarczmarek@astrouw.edu.pl (PK)},
G.~Wiktorowicz$^{1}$,
K.~I{\l}kiewicz$^{2}$,
R.~Smolec$^{2}$,
K.~St{\k e}pie{\'n}$^{1}$,
\newauthor
G.~Pietrzy{\'n}ski$^{2,3}$,
W.~Gieren$^{3,4}$,
K.~Belczynski$^{1}$
\\\\
$^{1}$Warsaw University Observatory, Al. Ujazdowskie 4, 00-478 Warsaw, Poland\\
$^{2}$Nicolaus Copernicus Astronomical Center of the Polish Academy of Sciences, Bartycka 18, 00-716 Warsaw, Poland	\\
$^{3}$Universidad de Concepci{\'o}n, Departamento de Astronom{\'i}a, Casilla 160-C, Concepci{\'o}n, Chile\\
$^{4}$Millenium Institute of Astrophysics, Santiago, Chile
}

\date{Accepted XXX. Received YYY; in original form ZZZ}

\pubyear{2016}

\begin{document}
\label{firstpage}
\pagerange{\pageref{firstpage}--\pageref{lastpage}}
\maketitle


\begin{abstract}
Single star evolution does not allow extremely low-mass stars to cross the classical instability strip (IS) during the Hubble time. {\color{black}However,} within binary evolution framework low-mass stars can appear inside the IS once the mass transfer (MT) is taken into account. Triggered by a discovery  of low-mass (0.26 M$_{\sun}$) RR Lyrae-like variable in a binary system, OGLE-BLG-RRLYR-02792, we investigate the occurrence of similar binary components in the IS, which set up a new class of low-mass pulsators. They are referred to as Binary Evolution Pulsators (BEPs) {\color{black}to underline} the interaction between components, which is crucial for substantial mass loss prior to the IS entrance. We simulate a population of 500\,000 metal-rich binaries and report that 28\,143 components of binary systems experience severe MT (loosing up to 90\% of mass), followed by at least one IS crossing in luminosity range of RR Lyrae (RRL) or Cepheid variables. A half of these systems enter the IS before the age of 4 Gyr. BEPs display {\color{black}a variety} of physical and orbital parameters, with the most important being the BEP mass in range $0.2-0.8$ M$_{\sun}$, and the orbital period in range $10-2\,500$ d. Based on the light curve only, BEPs can be misclassified as genuine classical pulsators, and as such they would contaminate genuine RRL and classical Cepheid variables at levels of 0.8\% and 5\%, respectively. We state that the majority of BEPs will remain undetected and we discuss relevant detection limitations.
\end{abstract}



\section{Introduction}
The number of pulsating variables detected in binary systems continuously grows. Binarity or multiplicity has been reported among various types of pulsators across the Hertzsprung-Russel Diagram (HRD) and at different stages of evolution, most of them being classical Cepheids \citep{szabados98}, $\delta$-Scuti stars \citep{soydugan10,liakos12}, pulsating sub-dwarfs \citep{lynas13}, and red giants \citep{gaulme14}. Pulsating variables in eclipsing binaries, in particular, constitute an invaluable laboratory to determine physical properties of stars, like masses and radii, with errors smaller than 1\% \citep[e.g.][]{pilecki13, gieren14}. Comparison of observational data against theoretical models of stellar structure and evolution leads to a considerably more detailed understanding of pulsational properties and evolutionary history of variable stars.

Standards candles, like classical Cepheids and RR Lyrae (RRL) stars, require the most profound examination as they are extensively used to determine distances and ages of galaxies and clusters. Unlike Cepehid binaries, RRL binaries have not been found yet. Although \citet{hajdu15} reported on 12 firm candidates for RRLs in binary systems, these findings need complementary spectroscopic observations and further analysis to be confirmed. A discovery of an RRL variable in eclipsing binary system would allow to measure with high accuracy physical parameters of this pulsator. Given high-quality observables, one could infer from pulsational models about other properties, like brightness \citep{bono03}, and reaffirm RRL status of a standard candle.

In course of the Araucaria Project, the basic astrophysical parameters of Cepheid variables in a number of eclipsing binary systems were established, with a dynamical mass being the most fundamental one \citep{pietrzynski10, pietrzynski11, pilecki15}. The same analysis was done in case of double-lined spectroscopic (SB2) eclipsing binary with a component exhibiting RRL-type pulsations, OGLE-BLG-RRLYR-02792 \citep[][hereinafter referred to as P12]{pietrzynski12}, but yielded surprisingly low mass of the pulsator, only 0.26\,M$_{\sun}$, which is clearly too small for genuine RR Lyrae star. 
{
\color{black} RR Lyrae stars are typically of mass $0.6 - 0.8$ M$_{\sun}$, which is sufficient to ignite helium in the $\sim 0.5$ M$_{\sun}$ degenerated core and radiate energy throughout the envelope built of zones of partial ionization in a way to generate and sustain pulsations. Although theoretical predictions allow for RRL mass as low as 0.36 M$_{\sun}$ \citep[provided massive envelope and efficient mass loss,][]{bono97},  no single star evolutionary scenario accounts for RR Lyrae star of mass only 0.26 M$_{\sun}$.
}
According to P12, OGLE-BLG-RRLYR-02792 emerged from the mass transfer (MT) episode, during which the initially more massive component removed most of its envelope and exposed degenerated helium core, placing itself in a narrow range of effective temperatures and luminosities characteristic for RR Lyrae instability strip. The MT was the key stage in this binary evolution, allowing to produce such a low-mass pulsator within the Hubble time. Thus, this puzzling object was named Binary Evolution Pulsator (BEP), highlighting the crucial role of the MT in its evolution. Together with the first BEP, new evolutionary channel producing RRL-like pulsators was recognized. Although the first BEP gained attention over the next couple years \citep{karczmarek12,smolec13} and its properties, evolution scheme, and detection methods were carefully studied, it is up to date the only known representative of its class.

BEP objects in non-eclipsing binary systems can be indistinguishable from RRL or Cepheid variables, given only the light curve. One must account for the possibility that some stars inside the classical instability strip (IS\footnote{Herein the IS refers to the classical instability strip defined by Eq. \ref{eq:IS} and within the luminosity range $10-10^5$\,L$_{\sun}$.}) may actually be unrecognized BEPs. This contamination value, i.e. the percentage of BEPs misclassified as genuine pulsators, may serve as an indicator of the expected number of BEPs to be discovered or re-classified among known RRL and Cepheid stars. Relatively large number of undetected BEPs may affect statistics-based calculations involving classical pulsators, like age or distance determinations.

The purpose of this work is to determine the number of binary systems that cross the IS, with the use of population synthesis code, StarTrack \citep{belczynski02,belczynski08}. We aim to characterize these systems in statistical form, with a particular attention to those physical parameters which may be established from observations, like masses, radii, and orbital periods. With this approach not only we calculate the contamination of RRL variables from BEPs and compare it with previous estimations from P12, but also we inspect BEP occurrence in the IS above the luminosity regime of RRL stars, in Cepheid domain. Our study relies on the assumption that all stars inside the IS pulsate. Few exceptions of non-pulsating stars inside the IS \citep{guzik14,murphy15} lie on the HRD well below the lower luminosity threshold set in our simulations. Additionally, we determine evolutionary channels through which BEPs are formed and their evolutionary fate as low-mas white dwarf (WD) stars.

Although the population synthesis of low-mass binaries was extensively used for analysis of WDs \citep[e.g.][]{iben93,driebe98,podsiadlowski02}, and resulted in evolutionary tracks that do cross the IS in RRL/Cepheid domain, the evolution inside the IS has not been paid much attention. The study of \citet{neilson15} targeted classical Cepheids and reported that majority of systems underwent the MT, but excluded them from further analysis. Our work complements existing population synthesis studies with this particular part of the binary evolution when one of components crosses the IS after substantial mass loss due to the interaction with the companion. We also challenge the evolutionary scheme of binary pulsators and claim that they can be found inside the IS even after the substantial mass loss.

\section{Binary calculations}
The objective of this section is to extract BEP objects from a synthetic population of metal-rich ($Z=0.02$) binaries. 
{
\color{black}
This particular metallicity was assumed by P12 in the study of the first BEP, which mimics RRL pulsations. Within the canonical framework of single star evolution, RRL variables reflect late evolutionary stages of low-mass and metal-poor stars, whereas low-mass and metal-rich stars fail to become RRL variables as they do not depart from the Red Giant Granch (RGB) enough to enter the IS. This view was recently undermined, as spectroscopic \citep{walker91, wallerstein12} and photometric \citep{pietrukowicz15} observations disclosed a few RRL stars with close-to-solar metallicities in the solar neighbourhood and in the Galactic Bulge. These findings were supported by the theoretical investigation of \citet{bono97, bono97a}, who proposed that metal-rich ($Z=0.02$) RRLs not only can exist, but also can be much younger ($\sim 0.9$\,Gyr) and of lower mass ($\sim 0.36$\,M$_{\sun}$) than previously thought, provided a substantial mass loss during the RGB phase. 
In a broader context of binary evolution, this mass loss might be attributed to the interaction with a companion.
}

Following P12, we define BEP objects as components of binary systems, which experienced the interaction with the companion in form of MT prior to IS entrance. We further extend this definition by a comparison to a single star of the same initial physical parameters as BEP progenitor, and we state that if the single star is predicted to cross the IS, then the binary component is defined as BEP, if it crosses the IS (after the interaction with the companion) at significantly different age than its single star analogue.

Above definition excludes stars which cross the IS regardless of the occurrence of MT episode in the past, and when they do so, their age is in agreement with stellar evolution scenarios for single stars.

\subsection{StarTrack population synthesis code}
\label{sec:calculations}
The StarTrack (ST) population synthesis code \citep{belczynski02,belczynski08} is based on revised formulae from \citet{hurley00,hurley02}, fitted to detailed single star models with convective core overshooting across the entire HRD \citep{pols98}. A number of enhancements implemented to the ST code account for wind accretion through Bondi-Hoyle mechanism, atmospheric Roche-lobe overflow \citep{ritter88}, wind Roche-lobe overflow \citep{mohamed12,abate13}.

We used ST to generate a population of 500\,000 binaries at zero-age main sequence (ZAMS) characterized by four parameters (initial mass of the more massive component, initial mass ratio, initial orbital separation, and initial eccentricity) randomly chosen from the following distributions:
\begin{itemize}
\item broken power law initial mass function \citep[IMF,][]{kroupa03} for the mass of the primary $M_\mathrm{A}$ in range $0.5 -  150.0$\,M$_{\sun}$,
\item flat distribution of mass ratio of secondary to primary $q=M_\mathrm{B}/M_\mathrm{A}$ \citep*{kobulnicky07} in range $q_\mathrm{min} - 1$, where $q_\mathrm{min}$ is the mass ratio that determines the lower mass limit of the secondary $M_\mathrm{B} = 0.08$\,M$_{\sun}$,
\item flat distribution of the logarithm of binary separation \citep{abt83} in range $a_\mathrm{min} - 10^5$ R$_{\sun}$, where $a_\mathrm{min}$ is the doubled sum of components' radii at periastron,
\item  thermal-equilibrium distribution of eccentricities \citep{heggie75} $\Xi(e) = 2e$ in range $0-0.99$.
\end{itemize}

Systems with orbital separation such that their orbital period is shorter than the threshold value $P_\mathrm{orb} = 4.3$ d, have initially circularized orbit and orbital rotation synchronized with the components' rotation due to pre-MS tidal circularization \citep{mathieu94}. The maximum evolutionary age considered for each binary is 14 Gyr. For symmetry reasons only binaries with $M_\mathrm{A} > M_\mathrm{B}$ are evolved.

Apart from these four parameters drawn for each system from above distributions, ST was run with parameters fixed throughout the entire population synthesis, among which the most important are: metallicity at ZAMS $Z = 0.02$, and fraction of mass lost from the system during the MT $f = 0.5$.

In order to make the ST code more suitable for our study of low-mass binaries, we implemented alternative formulae for rates of mass loss and angular momentum loss \citep*{stepien12}:
\begin{align}
\dot{M}_\mathrm{A,B} &= -10^{-11} R^2_\mathrm{A,B}\\
\dot{H}_{\rm orb} &= 
\begin{cases}
-4.9 \times 10^{41} (R_\mathrm{A}^2 M_\mathrm{A} + R_\mathrm{B}^2 M_\mathrm{B})/P_\mathrm{orb} & \mbox{if } P_\mathrm{orb} > 0.4\,\mathrm{d} \\[0.3em]
-4.9 \times 10^{41}(R_\mathrm{A}^2 M_\mathrm{A} + R_\mathrm{B}^2 M_\mathrm{B})/0.4 & \mbox{if } P_\mathrm{orb} \leq 0.4\,\mathrm{d}
\end{cases}
\label{eq:stepien}
\end{align}
Angular momentum is expressed in cgs units, period in days, masses and radii in solar units. The rates of angular momentum loss and mass loss are computed per time unit of a year. Both $\dot{H}_{\rm orb}$ and $\dot{M}_\mathrm{A,B}$ apply to stars with masses $M_\mathrm{A,B} \leq 1.5$\,M$_{\sun}$ that reached orbit and spin synchronization. The impact of these changes on the evolution of low-mass binaries is marginal, because of their limited range of action (only low-mass main sequence stars on narrow orbits).

Special attention was devoted to the refinement of the MT -- one of the most vulnerable processes in a binary evolution. In our study, at the onset of the MT the donor is always more massive component and thus the mass is transferred in thermal timescale until one of two possibilities occurs: (i) the gainer accretes the mass until the mass ratio reversal, wherefore the MT proceeds in the nuclear timescale, or (ii) the system develops the common envelope (CE), ejects it and emerges as a close WD binary or a merger. BEP is expected to be formed under the first scenario, provided the simulation from P12. To achieve a similar result, more liberal treatment of MT was favored, so that the CE scenario was executed only when the donor expanded beyond its outer Roche Lobe \citep*{pavlovskii15}. The approximated radius of the outer Roche Lobe \citep*{yakut05} was implemented into the ST code in a form:
\begin{equation}
R_\mathrm{L,outer} =
\begin{cases}
\dfrac{a\ (0.49 q^{2/3} + 0.15)}{0.6 q^{2/3} + \log(1 + q^{1/3})} & \mbox{if } q  \leq 1\\[1em]
\dfrac{a\ (0.49 q^{2/3} + 0.27 q - 0.12 q^{4/3})}{0.6 q^{2/3} + \log(1 + q^{1/3})} &  \mbox{if } q  > 1
\end{cases}
\label{eq:outerRL}
\end{equation}
where $a$ is a components' separation in solar radii and $q = M_\mathrm{donor} / M_\mathrm{gainer}$.

\subsection{Extraction of BEP objects}
Our study focuses on stars in binary systems, which experienced interaction with companions via MT and later entered the IS. Simplified red and blue edges of the IS are adapted from MESA code \citep[Modules for Experiments in Stellar Astrophysics,][]{paxton15} as follows:
\begin{equation}
\begin{array}{l}
\log T_\mathrm{red}~ = -0.05 \log L + 3.94\\[0.3em]
\log T_\mathrm{blue} = -0.05 \log L + 4.00
\end{array}
\label{eq:IS}
\end{equation}
where $L$ is stellar luminosity in range $10-10^5$\,L$_{\sun}$, and $T_\mathrm{red}$, $T_\mathrm{blue}$ are effective temperatures at red and blue edges of the IS, respectively. 
{
\color{black}For the sake of consistency of the results, only one pair of formulae for red and blue edges was used along the entire IS. Because synthesis population uses simplified formulae for stellar and binary evolution and yields results of only statistical importance, more detailed boundary formulae (i.e. metallicity dependent and designated separately for RR Lyrae stars and classical Cepheids) would add complexity to the calculations without a significant change in the results.
}

Let us recall, that BEP is defined not only as a component of a binary system, which enters the IS after the MT, but also as an object which due to the interaction with the companion via MT enters the IS at the age significantly different than its single stellar analogues, provided that initial physical parameters of both BEP progenitor and its single star analogue allow them to cross the IS {\color{black}within the single} stellar evolution framework. These two conditions are crucial for BEP occurrence, and are followed by a number of accompanying conditions, contained in a filtering algorithm described below. Every system from our synthetic population was subjected to this algorithm, and if all conditions were satisfied, the system was marked as BEP.

Only systems that experienced MT onset before core helium ignition were taken into account, however we accepted the possibility of helium ignition during MT phase or after its termination. The MT rate itself was not limited, but if the maximum MT rate was higher than $10^{-7}$\,M$_{\sun}$\,yr$^{-1}$ (the value of constant MT rate in the first BEP model, P12) for longer than 5\% of the total MT phase, the system was rejected. 
Only stars with substantial mass change (at least 20\% of initial mass) due to MT were accepted.
Large mass loss inside the IS induced by stellar winds (especially if the system crossed the IS towards high temperatures) might indicate the beginning of the planetary nebula phase. In order to exclude such instances, only BEP candidates with the mass difference of less than 5\% between the beginning and the end of the IS were accepted, and only if their mass loss rate was smaller than $5 \times 10^{-8}$\,M$_{\sun}$\,yr$^{-1}$ during the IS. 
The upper limit for BEP mass was chosen as to exclude massive objects, which could achieve the same evolutionary stage following only single star evolution. If a BEP candidate (after interaction with the companion via MT) was demonstrated to cross the IS at similar age as its single stellar analogue, the candidate was rejected. {\color{black}This assumption allows us to set} the upper limit of BEP mass to 0.9\,M$_{\sun}$. No lower mass limit was set.

The evolution across the IS was carefully monitored on the HRD in order to account for one of four cases: (i) a star crosses the IS blueward, (ii) a star crosses the IS redward, (iii) a star makes a loop inside the IS, or 
{
\color{black} (iv) a star appears inside the IS on the Horizontal Branch (HB) after the helium ignition in the core. These instances appear in more detail in the following section.
}

\section{Results}
Out of 500\,000 simulated binaries, we extracted a sample of 28\,143 BEP objects. Vast majority of our sample experiences only one IS crossing towards the blue edge of the IS, yet in a small sub-sample of 3.5\% of all systems, one component exhibits additional one or two IS crossings on the HRD. 
{
\color{black}Additional IS crossings take place when stars re-enter the IS while performing a Cepheid-like  ``blue loop'' or depart from the Asymptotic Giant Branch (AGB) towards higher effective temperatures, and will be elaborated in section \ref{sec:evol_routes}.
}
As we study each IS crossing independently, we recognize them all as separate phenomena in our sample. Therefore, the actual number of BEP sample, in sense of individual IS crossings, is 29\,386.

BEP objects present a variety of masses, luminosities, orbital periods, and evolutionary stages. Therefore we decided to introduce basic classification in terms of their evolutionary type inside the IS. 
{
\color{black} Three evolutionary types are identified and characterized below, whereas their evolutionary scenarios and quantitative description is given later in this section.
\begin{itemize}
\item \emph{RGB-core} type designates stars which lost a substantial part of the envelope before approaching the tip of the RGB, and as a consequence departed from the RGB before the helium ignition in the core. Therefore they cross the instability strip as almost entirely stripped helium cores, which structurally resemble cores of RGB single stars. A small fraction of RGB-core BEPs ignites helium after leaving the IS (in a post-BEP phase), and even smaller re-enters the IS after helium ignition as different BEP type.
\item \emph{HeB-core} type describes objects which experience helium burning in their cores while crossing the IS. A single star analogue for HeB-core BEP is a Horizontal Branch star, yet BEPs may not populate the exact locus of the Horizontal Branch (HB) on the HRD, as their core masses are more dispersed than in HB stars.
\item \emph{AGB-core} type describes objects which departed from the Asymptotic Giant Branch due to the mass transfer or/and significant mass loss via stellar winds. None of AGB-core BEPs commenced thermal pulses and so they resemble early post-AGB single stars \citep[e.g.][]{dcruz96,oconnell99}. They possess an envelope of such a low mass, that they are virtually stripped cores of AGB stars, traversing the IS. 
\end{itemize}
}

{
\color{black} This large and diverse BEP sample enabled
}
to construct a reliable statistics and distributions of a number of BEP features and, based on them, infer about crucial BEP characteristics. In Table \ref{tab:main_table} we present exemplary BEP physical and evolutionary parameters, while the full-length machine readable table is available online at \url{http://www.astrouw.edu.pl/~pkarczmarek/projects/beptable.dat}. 

\begin{table*}
\caption{Physical and orbital parameters of BEP objects averaged over IS time span, BEP progenitors at ZAMS and their companions. Companion's parameters are always captured either on ZAMS or while the other component is in BEP phase. Subsequent columns represent: mass of BEP progenitor at ZAMS, mass of a companion at ZAMS, orbital period at ZAMS, time of IS entrance (beginning of BEP phase), time spent inside the IS (duration of BEP phase), BEP mass, mass of BEP's companion, orbital period of BEP system, BEP radius, radius of BEP's companion, BEP luminosity, luminosity of BEP's companion, type of BEP, and type of BEP's companion.}
\label{tab:main_table}
\begin{tabular}{\decol \decol \decol \decol \decol \decol \decol \decol \decol \decol \decol \decol c c}
\hline
\noalign{\smallskip}
\multicolumn{1}{c}{$M_\mathrm{BEP}^\mathrm{ZAMS}$} & 
\multicolumn{1}{c}{$M_\mathrm{com}^\mathrm{ZAMS}$} & 
\multicolumn{1}{c}{$P_\mathrm{orb}^\mathrm{ZAMS}$} &
\multicolumn{1}{c}{$t^\mathrm{IS}$} &
\multicolumn{1}{c}{$\Delta t^\mathrm{IS}$} &
\multicolumn{1}{c}{$M_\mathrm{BEP}^\mathrm{IS}$} & 
\multicolumn{1}{c}{$M_\mathrm{com}^\mathrm{IS}$} & 
\multicolumn{1}{c}{$P_\mathrm{orb}^\mathrm{IS}$} &
\multicolumn{1}{c}{$R_\mathrm{BEP}^\mathrm{IS}$} & 
\multicolumn{1}{c}{$R_\mathrm{com}^\mathrm{IS}$} & 
\multicolumn{1}{c}{$L_\mathrm{BEP}^\mathrm{IS}$} & 
\multicolumn{1}{c}{$L_\mathrm{com}^\mathrm{IS}$} & 
\multicolumn{1}{c}{type$_\mathrm{BEP}^\mathrm{IS}$} &
\multicolumn{1}{c}{type$_\mathrm{com}^\mathrm{IS}$}
\\
\multicolumn{1}{c}{(M$_{\sun}$)} & 
\multicolumn{1}{c}{(M$_{\sun}$)} & 
\multicolumn{1}{c}{(d)} &
\multicolumn{1}{c}{(Myr)} &
\multicolumn{1}{c}{(kyr)} &
\multicolumn{1}{c}{(M$_{\sun}$)} & 
\multicolumn{1}{c}{(M$_{\sun}$)} & 
\multicolumn{1}{c}{(d)} &
\multicolumn{1}{c}{(R$_{\sun}$)} & 
\multicolumn{1}{c}{(R$_{\sun}$)} & 
\multicolumn{1}{c}{(L$_{\sun}$)} & 
\multicolumn{1}{c}{(L$_{\sun}$)} & 
\multicolumn{1}{c}{} &

\\
\hline
\noalign{\smallskip}
1.12 & 0.61 & 28.23 & 8445.58 & 226.18 & 0.28 & 0.93 & 54.50 & 6.02 & 0.84 & 97.92 & 0.52 & \mbox{RGBc} & \mbox{MS} \\
3.08 & 2.18 & 9.00 & 364.22 & 39.85 & 0.39 & 3.51 & 87.19 & 12.41 & 2.49 & 325.64 & 166.84 & \mbox{RGBc} & \mbox{MS} \\
1.28 & 0.73 & 200.20 & 5044.45 & 147.05 & 0.29 & 1.21 & 65.07 & 6.81 & 1.19 & 119.71 & 1.89 & \mbox{RGBc} & \mbox{MS} \\
3.52 & 2.45 & 89.04 & 272.24 & 5818.48 & 0.87 & 3.76 & 349.78 & 13.97 & 2.73 & 400.06 & 226.52 & \mbox{HeBc} & \mbox{MS} \\
1.93 & 1.70 & 774.75 & 1525.55 & 10.18 & 0.54 & 2.37 & 620.00 & 44.79 & 2.50 & 2770.41 & 42.93 & \mbox{AGBc} & \mbox{MS} \\
2.61 & 2.22 & 2225.84 & 557.45 & 38.16 & 0.39 & 3.32 & 102.63 & 13.85 & 2.70 & 390.84 & 149.06 & \mbox{RGBc} & \mbox{MS} \\
1.83 & 1.17 & 130.83 & 1561.88 & 13.70 & 0.36 & 1.90 & 209.43 & 15.77 & 1.69 & 486.51 & 13.81 & \mbox{RGBc} & \mbox{MS} \\
2.91 & 2.70 & 53.52 & 522.04 & 9.82 & 0.66 & 3.77 & 623.50 & 48.88 & 21.98 & 3192.20 & 235.46 & \mbox{AGBc} & \mbox{HeB} \\
\hline
\end{tabular}
\begin{flushleft}
Inside the IS, type of BEP and companion are denoted in the following way: MS -- main sequence star, RGBc -- RGB-core BEP, AGBc -- AGB-core BEP, HeBc -- HeB-core BEP, HeB --  core helium burning star.\\
This table is available in its entirety in a machine-readable format in the  website \url{http://www.astrouw.edu.pl/~pkarczmarek/projects/beptable.dat}
\end{flushleft}
\end{table*}

\subsection{Evolutionary routes}
\label{sec:evol_routes}
\begin{figure}
\includegraphics[width=\columnwidth]{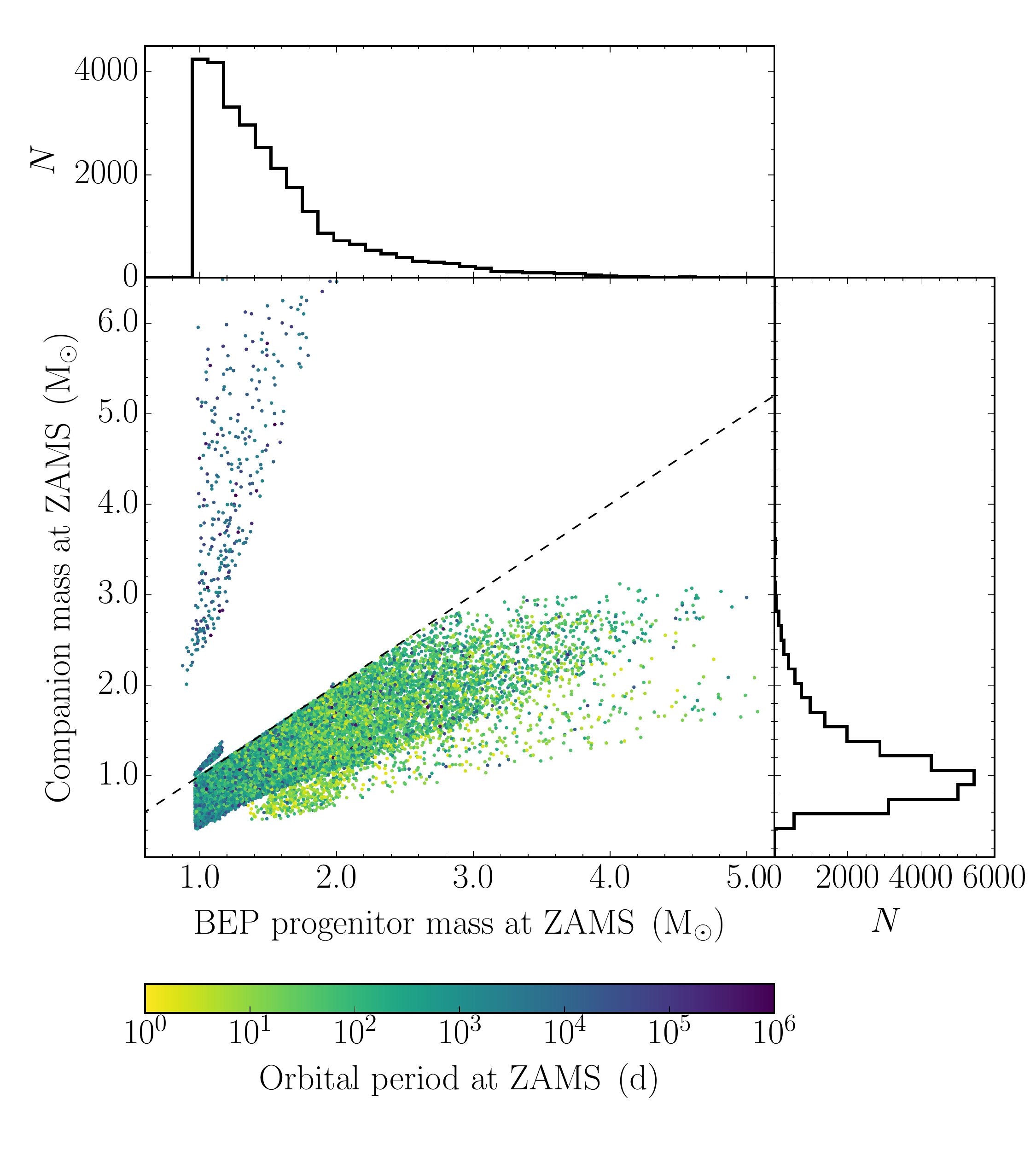}
\caption{Comparison of initial masses of BEP progenitor and its companion, together with initial orbital periods coded with colors. Dashed line determines the mass ratio 1.}
\label{fig:zams_mass_mass}
\end{figure}

All binaries hosting a BEP progenitor start an evolutionary route as two main sequence (MS) stars. Fig. \ref{fig:zams_mass_mass} presents comparison of initial masses of BEP progenitor and its companion together with their initial orbital periods. Masses of BEP progenitors cluster around 1.0\,M$_{\sun}$, and for most cases are larger than masses of their companions. Such initial mass configuration with BEP progenitor as the primary component determines general evolutionary route towards BEP formation, which includes single MT episode. However, in  1.1\% of our sample, BEP progenitors are less massive objects (initial secondaries) on wide initial orbits, and their peculiar evolution route towards BEP formation leads first through the CE, and then through the MT phase; in Fig. \ref{fig:zams_mass_mass} they form a cone above the diagonal line of mass ratio 1. 

\begin{figure}
\includegraphics[width=\columnwidth]{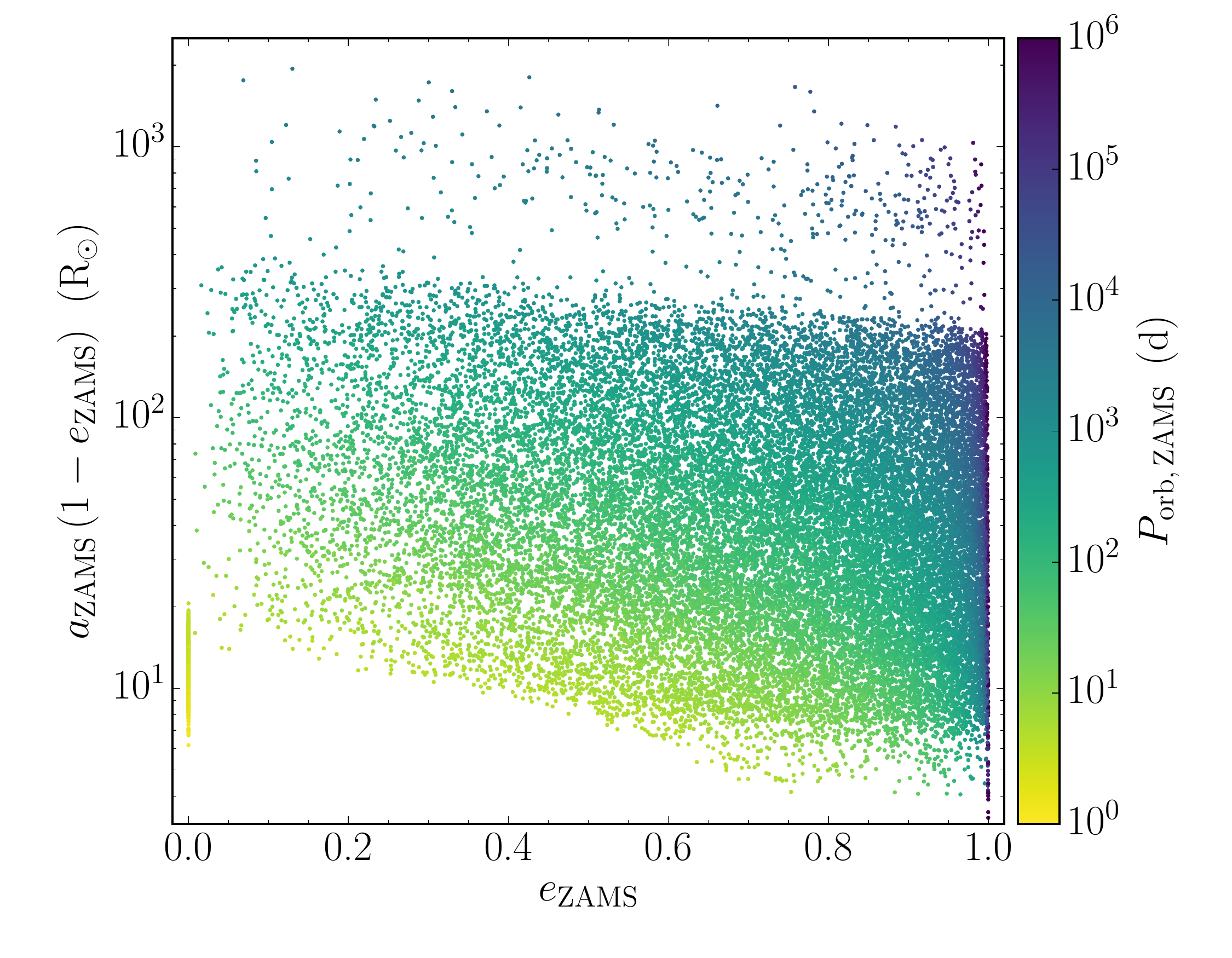}
\caption{Correlation between initial orbital separation and initial eccentricity of systems with BEP progenitor. Favorable combinations of orbital parameters fill a wide horizontal strip with upper and lower limits at approximately 5 and 300 R$_{\sun}$. Values of initial orbital period are color coded for reference.}
\label{fig:zams_e_a}
\end{figure}

Careful examination of initial orbital parameters of BEP progenitors in Fig. \ref{fig:zams_e_a} show, that BEPs form in systems of all eccentricities, and favorable eccentricity values depend mainly on the initial distribution of eccentricities. In case of thermal distribution of initial eccentricities adopted in our study, 50\% of all systems are found on highly eccentric ($e \geq 0.75$) orbits. The exception are few systems (2.8\%) on narrow orbits, that due to tidal circularization have eccentricity $e=0$. BEP progenitors reside inside a wide strip between two roughly constant values of y-axis, $5\,\mathrm{R}_{\sun} \leq a_\mathrm{ZAMS}(1-e_\mathrm{ZAMS}) \leq 300\,\mathrm{R}_{\sun}$. This means that the initial orbital separation depends on initial eccentricity in such a way that the larger the eccentricity, the wider the orbit. 

\begin{figure*}
\includegraphics[width=17cm]{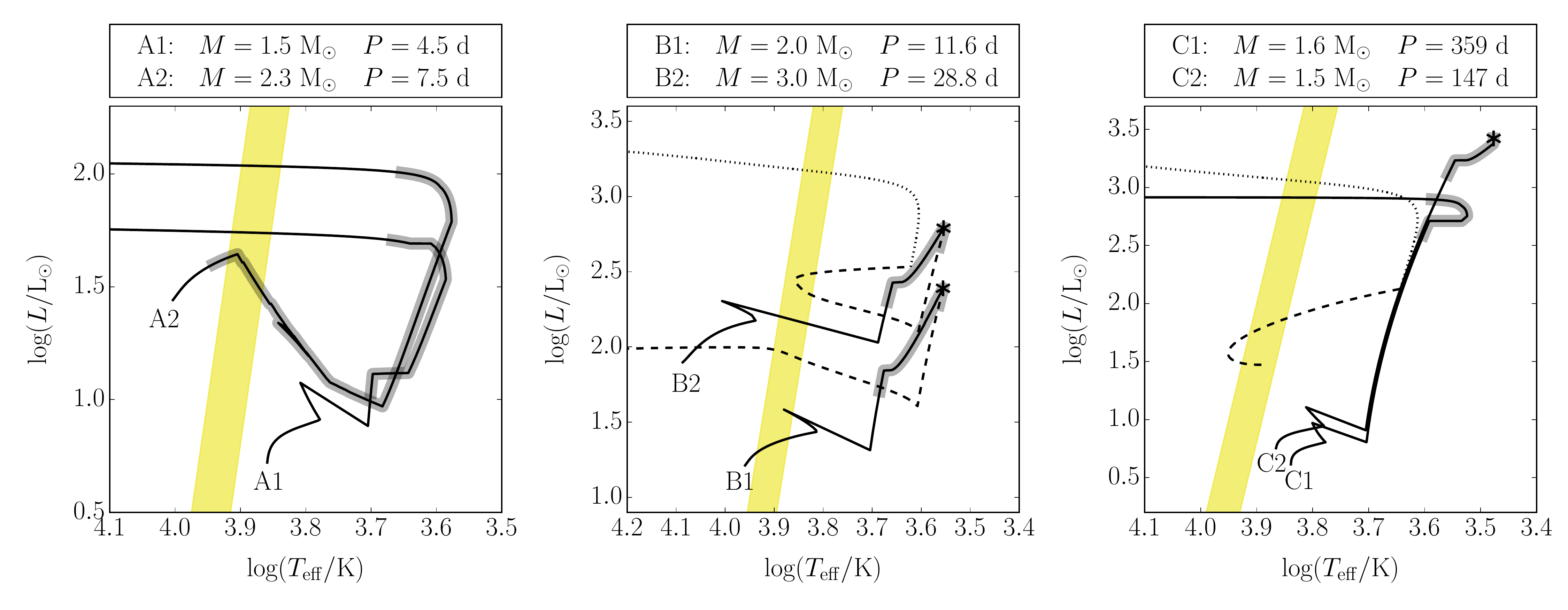}
\caption{Exemplary evolutionary routes of BEP objects on the Hertzsprung-Russel diagram, divided into three categories by the value of initial orbital period. Examples A1 and A2 have the shortest initial orbital period, while examples C1 and C2 have the longest. Companions' tracks are omitted for clarity. Parameters $M$ and $P$ above each plot describe initial mass of BEP progenitor and initial orbital period, respectively. Routes plotted with solid, dashed, and dotted lines mark evolutionary stages of pre-core-helium burning, core-helium burning, and post-core-helium burning phases, respectively. Asterisks denote core helium ignition. Shaded parts on tracks indicate mass transfer phases. Vertical strips denote location of the classical instability strip.}
\label{fig:hrd_tracks}
\end{figure*}

In order to illustrate different evolutionary paths that lead to the formation of BEPs, we chose several representative examples from our synthetic population and showed them in Fig. \ref{fig:hrd_tracks}. Each evolutionary track presents evolution of BEP progenitor on the HRD (companion omitted for clarity), and is marked with a letter and an order number. In all examples BEP progenitors are $1.1 - 1.8$ times more massive as their companions. Shaded parts on tracks indicate MT phases. Tracks from group $A$ show evolution of BEP progenitor on initially narrow orbit ($P_\mathrm{orb} < 10$\,d). All BEP progenitors {\color{black}belonging to this group} experience MT relatively early, while still on the MS, in the Hertzsprung Gap (HG) or at the base of the RGB. They finish MT as they depart from the RGB and move towards high effective temperatures on the HRD. They cross the IS once. In this evolution scheme BEPs of RGB-core type are formed. The example $A1$ is the most similar to the evolutionary track from P12. In the example $A2$, BEP progenitor of initial mass 2.3\,M$_{\sun}$ first crosses the IS while still in HG and before the MT is terminated. As such, we do not recognize the star as BEP in this phase.

Group $B$ represents systems with intermediate initial periods (10\,d $< P_\mathrm{orb} <$ 100\,d), that start the MT when BEP progenitors ascend the RGB. The MT ceases just after the helium ignition in the core
{
\color{black} and a considerable decrease in the radius of BEP progenitor, which allows the star to preserve more massive envelope than in examples of $A$ group.
}
In this evolution scheme HeB-core type of BEP is formed when a star crosses the IS while burning helium in the core (dashed line). While in example $B1$ a star moves directly towards high effective temperatures (as examples in $A$ group), a star from example $B2$ makes a Cepheid-like 
{
\color{black}``blue loop''
}
 inside the IS, and enters the IS again after the helium exhaustion in the core (dotted line) as an AGB-core object. Example $B2$ represents a large family of objects with loop-like behavior: objects with more pronounced loops exit and re-enter the IS while still burning helium in the core; objects with smaller loops do not enter the IS until they depart from the AGB.

Examples from group $C$ present evolutionary tracks of BEP progenitors on wide initial orbits ($P_\mathrm{orb} >$ 100\,d). These stars start the MT late on the RGB, which leads to two main outcomes: either the BEP progenitor departs from the RGB without the helium ignition and crosses the IS as a star with RGB-core (example $C1$), or the MT ends at the onset of helium burning in the core, and BEP traverses the IS as a star with HeB-core ($C2$, dashed line) and again later as AGB-core object ($C2$, dotted line). A few variations of the example $C2$ are possible, as the star at the onset of core helium burning phase may appear outside or inside the IS, and make larger or smaller turn towards the AGB.

Let us mention systems with BEP progenitor's companion as primary, initially more massive component. These systems exist only on wide orbits ($P_\mathrm{orb} \gtrsim$ 2000\,d) and experience the CE episode, triggered by the primary, while the secondary (BEP progenitor) is sill on the MS, which leads to a considerable loss of mass of the primary. After the CE episode, the BEP progenitor is the more massive component and becomes a donor in the following MT episode. The evolutionary track of BEP progenitor in this scheme is similar to the example $A1$ and leads to formation of RGB-core BEP.

During the MT, BEP progenitors lose remarkably $40-90$\% of their initial masses, with the most frequent value of 75\%. It is noteworthy, that the most mass ($\sim$\,90\%) is lost by massive BEP progenitors in systems on narrow orbits, and the least ($\sim$\,40\%) by low-mass BEP progenitors on wide orbits.

\begin{figure}
\includegraphics[width=\columnwidth]{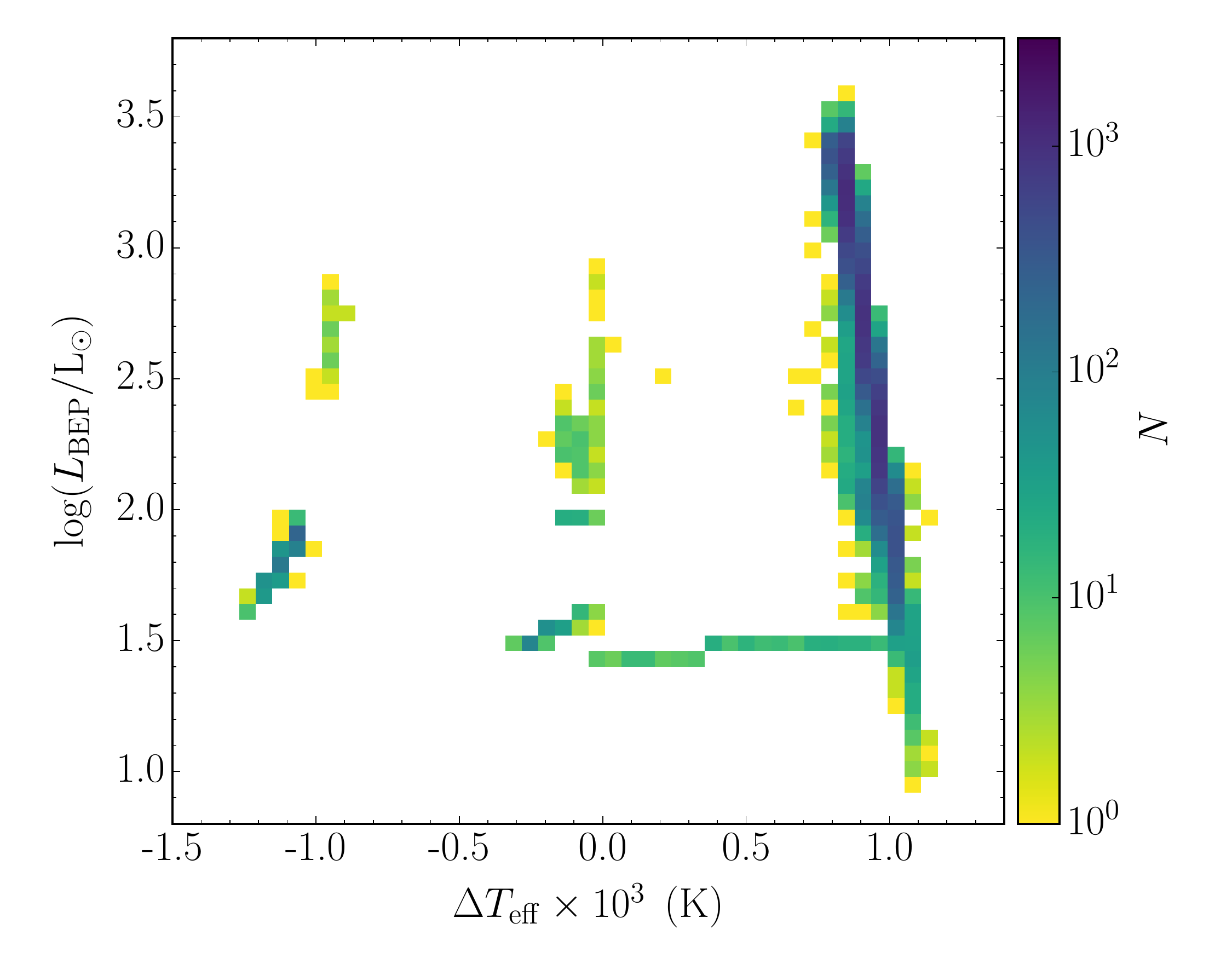}
\caption{Effective temperature difference of BEPs at the time of the entrance and exit from the instability strip. Colors code BEPs counts in each cell. Data cluster at temperatures of $-1000$ K, 0 K, and $+1000$ K, which correspond to stars moving redward, making a loop, and moving blueward, respectively.}
\label{fig:dteff-lum}
\end{figure}

Fig. \ref{fig:dteff-lum} presents a variety of IS crossings in our sample, in a form of a difference between effective temperatures at times of entrance and exit from the IS, confronted with luminosities averaged over IS time span. Three trends are visible as the data cluster around temperatures $-1000$\,K, 0\,K, and $+1000$\,K. Positive temperature difference corresponds to BEP objects moving blueward inside the IS, negative -- to objects moving redward, and the temperature close to 0 K denotes objects making a loop.
The horizontal line spanning from 0 to 1000 K at the $\log L \approx 1.5$\,L$_{\sun}$ denotes stars 
{
\color{black} which appeared inside the IS on the HB just after the helium flash, yet in various locations across the IS width. Population synthesis codes are not equipped with numerical routines allowing for detailed evolution through the helium flash, and thus the transition from the tip of the RGB to the HB is carried out in one time-step. This discontinuity is reflected in the evolutionary tracks of stars experiencing the helium flash as they re-appear suddenly in various locations across the HB, and possibly inside the IS.
}

The dispersion in the results is caused by the fact that systems were not captured at the exact point of entrance and exit of the IS, but at the point which in the simulations was the closest to IS edges.
It is noteworthy that RGB-core and AGB-core types occupy only the vertical space around $+1000$\,K, meaning they all cross the IS blueward, while HeB-core objects are found in all three vertical regions.

To sum up, BEPs appear in the IS as RGB-core, HeB-core, or AGB-core stars, and may cross the IS multiple times in one star's evolution, although the percentage of multiple-crossers is small (3.5\%). These three types constitute the entire BEP sample in following quantities: 82.4\% for RGB-core, 9.5\% for HeB-core, 8.1\% for AGB-core. 

{
\color{black} Although the post-BEP evolution is not the main focus of this paper, a brief discussion of post-BEP phases can be helpful in creating a complete and coherent picture of all evolutionary stages that concern BEP objects. In general, HeB-core and AGB-core BEPs evolve into carbon-oxygen white dwarfs (COWDs). Vast majority of RGB-core BEPs (96.3\%) evolve into so called \textit{flash-manqu\'{e}} stars, as they never ignite helium till the end of their evolution, and become helium white dwarfs (HeWDs). Small fraction (3.7\%) of RGB-core sub-sample ignites helium just after exiting the IS (0.07\% of which re-enters the IS as HeB-core BEPs) and ends as COWDs. Few RGB-core and HeB-core objects in their post-BEP evolution may become hybrid WDs (HybWDs), which essentially are white dwarfs with carbon-oxygen-helium cores and thin helium envelopes\footnote{From the evolutionary point of view HybWDs should be called COWDs since they burned helium in the core, but from the observational point of view they resemble HeWDs.}. 
Post-BEP objects, which used to be RGB-core and AGB-core types, become WDs in less than 2 Myr, and most commonly in $\sim 400$ kyr. Post-BEP evolution of former HeB-core objects towards the WD cooling sequence is much longer and can take up to 300 Myr, although the majority of systems become WDs in about $4-8$ Myr. During this time post-BEPs traverse the HRD in more less straightforward way until they reach the maximum effective temperature, and enter the WD cooling sequence. This maximum temperature  is spread over a wide range of $\log(T_\mathrm{eff}/\mathrm{K}) = 4.4 - 5.4$, and is usually lower for low-mass stars than for more massive ones.
The masses of post-BEP objects in their final WD stage do not differ substantially from the masses of BEP objects inside the IS, mostly because stars lose at most the rest of already minuscule envelope. Therefore the WD masses range from 0.2 to 0.7 M$_{\sun}$ with the average mass of 0.38 M$_{\sun}$. This average mass is dominated by low masses of HeWDs, and if only a sub-sample of HybWDs and COWDs was taken into account, the average mass would be 0.51 M$_{\sun}$.
Presented scheme resembles to some extent investigations of \citet{han00} and \citet{pradamoroni09}, who inspected the impact of binary interactions on the occurrence of low-mass WDs, and concluded that the mass transfer or the common envelope episode is most likely to form low-mass ($0.30 - 0.46$ M$_{\sun}$) WDs.
Discovery of WDs of even lower mass \citep[so called \emph{extremely low-mass} WDs,][]{hermes13} and pre-WDs \citep{maxted11,maxted14a} affirms this theory, especially since a considerable number of these stars is proved to be binary \citep{brown11}. 
A potential linkage between post-BEP stars and low-mass and extremely low-mass pre-WDs and WDs has not been establish yet, and can be an interesting subject for future investigation.
}

The final evolutionary stage in majority of our sample (98.6\%) is white dwarf (WD), with the following subgroups: helium WD (HeWD, 79.2\%), carbon-oxygen WD (COWD, 16.7\%), and hybrid WD (HybWD, 2.7\%). Few objects (0.8\%) end their evolution as helium stars 
{
\color{black}(exposed helium-burning cores emerged from enhanced mass loss or/and mass transfer while in the Hertzsprung gap or on the RGB), and they would proceed to COWD or HybWD stage,
}
if the simulation was run beyond 14 Gyr. The end evolutionary stage of the remaining 0.6\% stars of our synthetic population is either a massless remnant after supernova explosion or a neutron star; due to small number of these miscellanies, they are regarded as statistically unimportant and omitted in further analysis.

Over 60 various BEP evolutionary routes can be recognized, often differing only in details (e.g. a number of IS crossings during core helium burning phase, which gives a clue about the depth of the loop). For clarity reasons, only general and concise BEP evolutionary schemes are presented in Table \ref{tab:evol_routes}, together with the final stages of post-BEP evolution. 

\begin{table}
\caption{Most frequent BEP evolutionary routes.}
\label{tab:evol_routes}
\centering
\begin{tabular}{l r}
\hline
\noalign{\smallskip}
Evolution scheme$^{a}$ & \% \\
\hline
\noalign{\smallskip}
MT ~ \textbf{IS} ~ WD & 43.5 \\
MT ~ \textbf{IS} ~ CE ~ WD & 29.1 \\
MT ~ HeI ~ \textbf{IS} ~ CE ~ WD & 6.1 \\
MT ~ HeI ~ \textbf{IS} ~ WD & 4.1 \\
MT ~ \textbf{IS} ~ MT ~ CE ~ WD & 3.7 \\
MT ~ HeI ~ MT ~ \textbf{IS} ~ CE ~ WD & 3.2 \\
MT ~ \textbf{IS} ~ HeI ~ MT ~ CE ~ WD & 2.3 \\
MT ~ \textbf{IS} ~ MT ~ WD & 1.5 \\
CE ~ MT ~ \textbf{IS} ~ WD & 1.0 \\
MT ~ \textbf{IS} ~ HeI ~ \textbf{IS} ~ WD & 0.7 \\
MT ~ HeI ~ \textbf{IS} ~ SN ~ NS & 0.6 \\
Other & 4.2 \\
\noalign{\smallskip}
\hline
\noalign{\smallskip}
Pre-BEP evolutionary stages$^{b}$ & \% \\
\hline
\noalign{\smallskip}
MT(B) & 68.1 \\
MT(A) & 8.7 \\
MT(B) ~ HeI & 7.0 \\
MT(AB) & 4.9 \\
MT(B) ~ HeI ~ MT(C) & 3.2 \\
MT(A) ~ HeI & 2.5 \\ 
MT(AB) ~ HeI & 2.1 \\
CE(MS+AGB, MS+WD) ~ MT(B) & 1.0 \\
Other & 2.5\\
\noalign{\smallskip}
\hline
\noalign{\smallskip}
BEP + companion types$^{c}$ & \% \\
\hline
\noalign{\smallskip}
RGB-core + MS & 78.5 \\
HeB-core + MS & 8.9 \\
AGB-core + MS & 6.8 \\
RGB-core + RGB & 2.0 \\
RGB-core + WD & 1.8 \\
AGB-core + HeB & 0.9 \\
HeB-core + RGB  & 0.2 \\
HeB-core + HeB & 0.2 \\
Other & 0.7 \\
\hline
\noalign{\smallskip}
Fate$^{d}$ & \% \\
\hline
\noalign{\smallskip}
HeWD & 79.2 \\
COWD & 16.7 \\
HybWD & 2.7 \\
HeStar & 0.8 \\
Other & 0.6 \\
\hline
\multicolumn{2}{p{\columnwidth}}{$^a$All stars begin their evolution as main sequence (MS) stars and may go through following stages: MT -- mass transfer, CE -- common envelope, HeI -- helium ignition, IS -- instability strip, and finish evolutionary route as one of following types: WD -- white dwarf, NS -- neutron star.}\\
\multicolumn{2}{p{\columnwidth}}{$^b$MT stages are divided into types: A, AB, B, and C accordingly to the evolutionary stage of the donor at the MT onset: MS, HG/early RGB, RGB, and AGB, respectively. CE episode is detailed by the evolutionary types of both components prior and after the CE, given in the parentheses.}\\
\multicolumn{2}{p{\columnwidth}}{$^c$Types of BEP objects and BEP companions: MS -- main sequence star, RGB -- red giant branch star, HeB -- star burning helium in the core, WD -- white dwarf.}\\
\multicolumn{2}{p{\columnwidth}}{$^d$Evolutionary types of post-BEP objects at 14 Gyr: HeWD -- helium white dwarf, COWD -- carbon-oxygen white dwarf, HybWD -- hybrid white dwarf, HeStar -- naked helium star.}
\end{tabular}
\end{table}

\subsection{BEP properties inside the instability strip}
Almost all BEPs enter the IS just after the MT has stopped, over 80\% of entire BEP sample enters the IS up to 2 Myr since the MT stopped, for over 50\% this time is shorter than 0.35 Myr with the most frequent time of 0.07 Myr.

During the MT phase almost entire envelope is peeled off of BEP progenitor. 
{
\color{black} BEPs of type RGB-core possess envelope of mass $0.0046-0.0056$\,M$_{\sun}$ which is about $0.7-2.3$\% of their total masses. These values correspond to the upper limit of the range $0.001-0.005$\,M$_{\sun}$ reported by \citet{maxted14a} and agree with percentage of envelope mass employed by \citet{smolec13} in BEP pulsation models. AGB-core BEPs posses even smaller envelopes of mass tightly clustered between 0.0050 and 0.0051 M$_{\sun}$. Finally, BEPs with HeB-core possess the most diverse envelopes ($0.005 - 0.08$ M$_{\sun}$), which constitute $1-15\%$ of their total masses. This wide range of envelope masses result from the core helium ignition during the MT phase, which causes the stellar radius to decrease substantially, followed by the MT termination, and allows the star to preserve larger fraction of the envelope.
}

\begin{figure}
\includegraphics[width=\columnwidth]{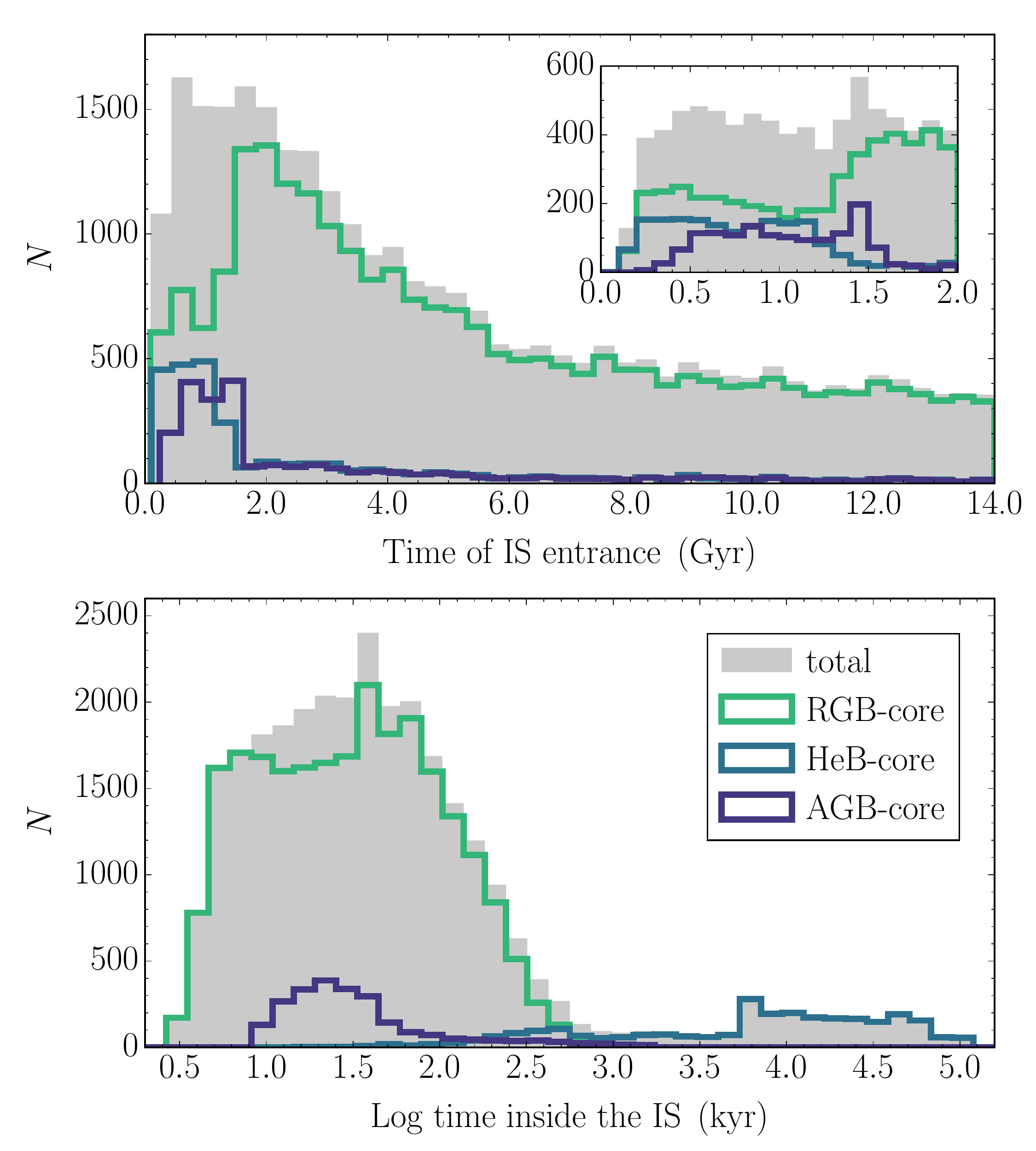}
\caption{Upper panel: distribution of BEP age at the IS entrance, the time range $0 - 2$ Gyr is magnified for clarity. Lower panel: duration of BEP phase. In both panels colors code BEP types.}
\label{fig:is_tenter_tdur}
\end{figure}

Fig. \ref{fig:is_tenter_tdur} presents the age of BEP objects when they enter the IS, and the time spent inside the IS, i. e. the duration of BEP stage. About 50\% of all IS crossings happen for stars younger than 4 Gyr, with the most frequent values between $0.5-2.5$ Gyr. Beyond the age of 6 Gyr the frequency of IS entrances is lower and stable until the end of the simulation at 14 Gyr. This stable trend reflects systems on initially wide orbits and with low-mass primaries, which start the MT relatively late.

The duration of BEP stage is determined by the width of the IS and by the stars' speed along evolutionary tracks on the HRD. This speed is determined by BEP evolutionary stage: RGB-core and AGB-core types move with the highest speed, while HeB-core type moves slowly due to core helium burning. Even so, the HeB-core type shows wide range of IS duration times, which is caused by various behaviors inside the IS and in the IS proximity (e.g. if a small part of a star's loop was inside the IS, BEP stage would last shortly). For our adopted constant width of the IS equal to 1300 K, the time spent inside the IS for RGB-core type ranges $3 - 500$ kyr with the wide plateau from 5 to 50 kyr, for AGB-core type IS duration clusters around 20 kyr, and for HeB-core type spreads $100\,\mathrm{kyr} \,-\, 120\,\mathrm{Myr}$ with the median time span at about 400 kyr. The second most frequent IS duration for HeB-core type of BEP is about 6.1 Myr and starts a plateau in range $\log \Delta t = 3.8-4.8$, though this frequent occurrence of HeB-core BEPs on the plateau is deceptive and results from adopted log scale. 

\begin{figure*}
\includegraphics[width=17cm]{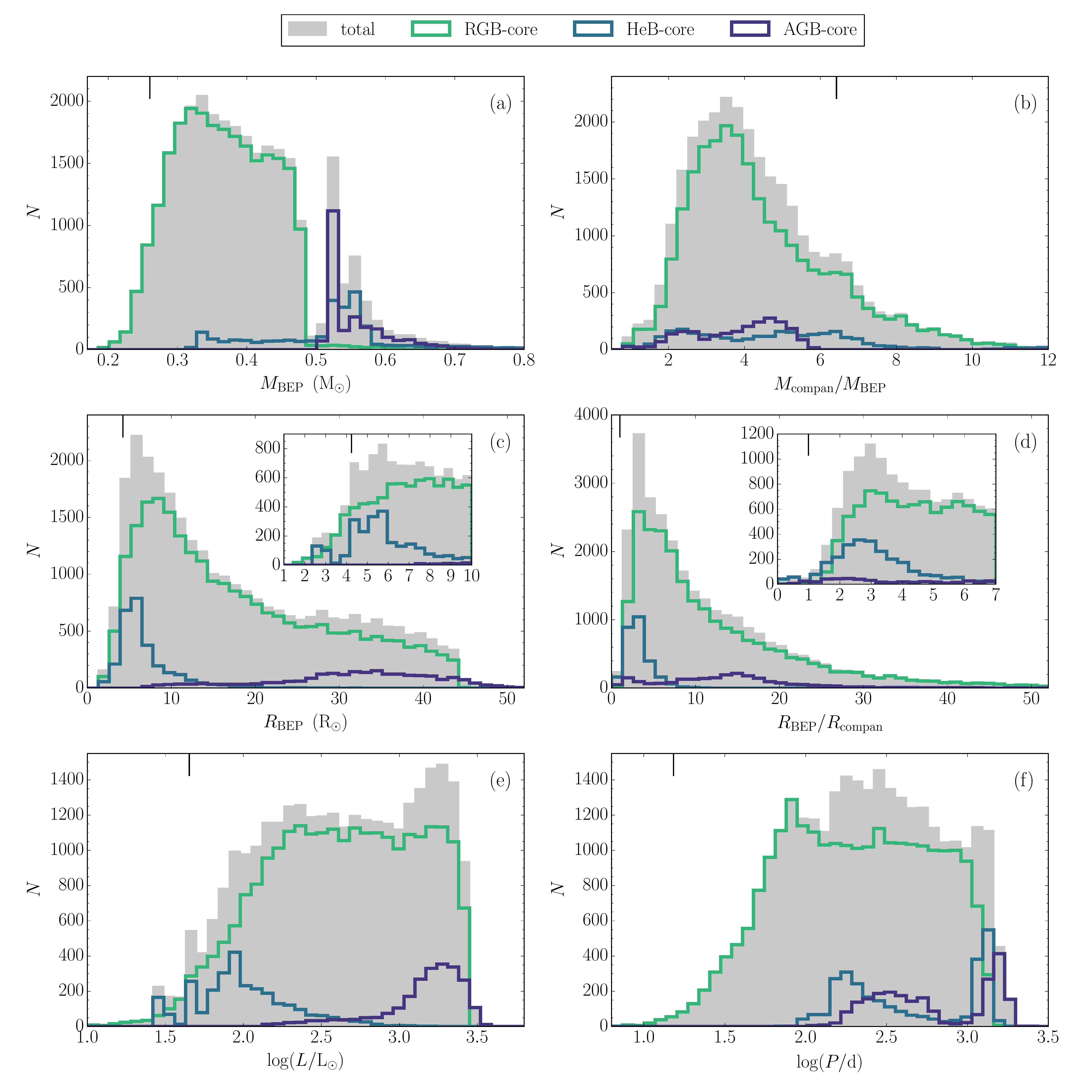}
\caption{Distributions of essential BEP observables with color coding for BEP types. All parameters are averaged over the IS time span. Subplots (c) and (d) in selected ranges are magnified for clarity. Black vertical line segment at the top of each subplot marks the value of the corresponding parameter observed in OGLE-BLG-RRL-02792.}
\label{fig:is_properties}
\end{figure*}

Essential BEP observables averaged over the IS time span are presented in Fig. \ref{fig:is_properties}. The distribution of BEP masses shows two peaks, at 0.52\,M$_{\sun}$ and 0.55\,M$_{\sun}$ for AGB-core and HeB-core types, respectively, and a broad plateau in range $0.28-0.48$\,M$_{\sun}$ (RGB-core type). 
{
\color{black} The core masses of RGB-core and AGB-core BEPs are virtually the same as their total masses, with the contribution from the envelope of only $\sim 1\%$ of total mass. HeB-core BEPs, in turn, have cores which constitute $85 - 99\%$ of their total mass. AGB-core BEPs have cores composed of helium, carbon and oxygen in proportions that reflect their evolutionary advancement at the time of departure from the AGB. The mass of carbon-oxygen cores ranges $55-95\%$ of the total core masses, with the median of 82\%. 
}
Overall, the total mass distribution confirms that majority of BEPs is expected to be extremely low-mass stars. For BEP masses below 0.35\,M$_{\sun}$, the most common mass ratio $M_\mathrm{compan}/M_\mathrm{BEP}$ is about 7, while objects above 0.35\,M$_{\sun}$ have a companion with mass such that the mass ratio is about $3-4$.

The distribution of radii depends on the evolutionary type of BEP, so that the AGB-core type, having higher luminosity, is expected to have also larger radius, as opposed to HeB-core type with much smaller radius. RGB-core type shows wide range of radii, from about 2\,R$_{\sun}$ to almost 45\,R$_{\sun}$, with the most frequent radius in range $6-10$\,R$_{\sun}$. The collateral distribution of BEP surface gravity $\log (g)$, derived from mass and radius, exhibits large diversity, ranging from 0.8 to 3.5. While for AGB-core type $\log (g)$ clusters around 1.1, and for HeB-core type -- around 1.5 and 2.6 (adequately to the two radii peaks at $\sim$\,2.8 and $\sim$\,5.0 R$_{\sun}$), $\log (g)$ for RGB-core type span over almost entire range, between 1.0 and 2.0.
The distribution of ratios of BEP radius to the companion radius extends up to 50, which reflects the variety of companions' evolutionary phases at a time of BEP crossing the IS. Vast majority of systems have radii ratio less than 6 (except for AGB-core type), and the median value is about 3.

The most frequent orbital period in our BEP sample is about 70 d. The entire contribution to this value comes solely from RGB-core type of BEP. The shortest orbital periods belong also only to RGB-core type and start at $4\ \mathrm{d}$. Moreover, objects with the shortest periods have also the lowest mass. Remaining types, HeB-core and AGB-core, have longer orbital periods, in range $\log (P/\mathrm{d}) = 2 - 3.3$. All orbits are circular, as a consequence of circularization and synchronization that occurred at the onset of the MT.

The distribution of BEP luminosities spreads from 10\,L$_{\sun}$ up to 3\,200\,L$_{\sun}$. AGB-core and HeB-core types show different distributions with maxima at $\log(L/\mathrm{L_{\sun}}) = 3.2$ and $\log(L/\mathrm{L_{\sun}}) = 1.9$, respectively. Luminosity of RGB-core type is distributed approximately log-uniformly within the range $2.2 - 3.4$. 
Noteworthy, luminosity range $\log (L/\mathrm{L_{\sun}}) = 1-2$ corresponds to the luminosity regime of RR Lyrae variables, and luminosities $\log (L/\mathrm{L_{\sun}}) > 2$ are characteristic for classical, anomalous, and type II Cepheids.

{
\color{black} Period-luminosity relation is simple yet one of the most powerful characteristics of pulsators in the IS, including RRL and Cepheid variables. Based on this relation and the distribution of periods of a given class of variables, one can infer about average luminosities of these stars, and further about their physical parameters. Indeed, the general information about the internal structure of a star and its pulsational period is encoded in the formula of \citet{ritter79}, $Q = P \sqrt{\rho/\rho_{\sun} }$, where the pulsation constant $Q$ varies among different types of variables. Above operation can be reversed to calculate pulsational periods, if pulsation constant $Q$ and physical parameters of variables are known. Precise measurements of the first BEP's mass, radius, and pulsation period, carried out by P12, yields $Q_\mathrm{BEP} =  0.0366 \pm 0.0033$\,d. This value is indistinguishable from the pulsational constant of RRL stars, $0.032 \lesssim Q_\mathrm{RRL} \lesssim 0.04$ d \citep[e.g.][]{christy66, jorgensen67}. In fact, the $Q$ value in range $0.03 - 0.04$ is a general characteristic of stars in the classical instability strip \citep{saio98}. However, propagation of $Q_\mathrm{BEP}$ from just one instance of BEP on the entire BEP sample is uncertain and dangerous practice, and the result need not to be correct. The link between the luminosity and the pulsational period of BEPs remains unknown and cannot be claimed to be the same as in classical pulsators until statistically relevant number of BEPs is discovered (for further discussion see section \ref{sec:detection}).
}

\section{Discussion}
\subsection{Comparison with observations}
So far, only one BEP has been discovered \citep{soszynski11} and studied \citep[P12,][]{karczmarek12,smolec13}. This first BEP, OGLE-BLG-RRLYR-02792, is an extremely-low mass (0.26\,M$_{\sun}$) component of an eclipsing binary system, with 1.67\,M$_{\sun}$ companion, and on a wide orbit of 32.2\,R$_{\sun}$, which corresponds to the orbital period $P_\mathrm{orb} = 15.24$\,d. In contrast to large mass ratio $M_\mathrm{compan}/M_\mathrm{BEP} = 6.42$, radii of components are virtually the same and equal to 4.24\,R$_{\sun}$ and 4.27\,R$_{\sun}$ for the BEP and the companion, respectively.

In order to compare BEP evolutionary scheme presented by P12 with our results, we ran an individual simulation with the ST code specifically for a binary with initial masses 1.4 and 0.8\,M$_{\sun}$ and initial orbital period 2.9\,d.
Our ST simulation leads to the MT after only 3.5\,Gyr from ZAMS, while the MT onset in the simulation of P12 starts at the age of 5.4\,Gyr. Our calculations predict that BEP progenitor loses 83\% of its initial mass during the MT, and enters the IS at the luminosity of 28\,L$_{\sun}$. The calculations of P12 agree as for the percentage of mass lost during the MT (81\%) but predict the luminosity of BEP to be slightly higher, i.e. 45\,L$_{\sun}$. More discrepancies are found inside the IS, as our calculated BEP mass (0.226\,M$_{\sun}$) and radius (2.79\,R$_{\sun}$), as well as companion's mass (1.38\,M$_{\sun}$) and radius (1.42\,R$_{\sun}$), and the orbital period (13.56\,d) are significantly smaller than the results of P12.

The parameters of the first BEP are the most similar to the system from our synthetic population with components' initial masses of 1.66\,M$_{\sun}$ and 1.13\,M$_{\sun}$, and the initial orbital period of 5.8\,d. During BEP stage, components have masses 0.26\,M$_{\sun}$ (BEP) and 1.65\,M$_{\sun}$ (companion), radii 4.7\,M$_{\sun}$ (BEP) and 1.6\,M$_{\sun}$ (companion), and the orbital period of 34.0\,d. Comparison of this synthetic system with observed BEP parameters leads to the conclusion, that the results of P12 and ours show systematic dissimilarities, which are most likely caused by the differences in employed codes, among which the most noticeable is treatment of the MT (P12 assumed strictly conservative MT, we used non-conservative MT with half of mass lost by donor transferred to the companion).

Let us highlight here, that synthesis population codes have to handle a large number of parameters, some of which not only are poorly calibrated from scarce observational data but also are represented by wide range of values yielded from detailed simulations. Uncertainties are related, among others, to the rate of angular momentum loss during the MT, the conservation of the MT, the length of time-steps, the efficiency of orbital energy and angular momentum loss in CE ejection. These uncertainties are acknowledged, but cannot be properly addressed until better observational constraints are provided.

Notwithstanding, all BEP observables derived by P12 fit into the distributions presented in Fig. \ref{fig:is_properties}, which confirms, that even though the exact evolutionary path of an individual binary is not precisely reproduced by the ST code, the statistical properties of the simulated population of BEPs well agree with observed BEP properties.

Long orbital periods of BEPs from our synthetic population can be confronted with periodical changes due to light-travel time effect found in the light curves of 12 candidates for RR Lyrae binary stars reported by \citet{hajdu15}. Logarithmic periods $\log (P_\mathrm{orb}/\mathrm{d})$ of these additional changes range from 3.09 to 3.66, and only slightly overlap our logarithmic period distribution (Fig. \ref{fig:is_properties}f). Thus, we expect these candidates to be genuine RR Lyrae binaries, once their binarity is confirmed. 

{
\color{black} Binarity among type II Cepheids is more common, and the number of confirmed type II Cepheids in binary systems continuously grows \citep[e.g.][]{harris89,soszynski10}. Type II Cepheids are low mass stars \citep[$0.52-0.59$ M$_{\sun}$,][]{bono97c} with helium-exhausted cores and luminosities in the wide range $\log (L/\mathrm{L_{\sun}}) \approx 2.0-4.0$. The same features are present in our sample of AGB-core BEPs. Another similarity are the orbital periods of our AGB-core BEPs of approximately $120-2500$ d, and orbital periods of binaries hosting type II Cepheids, which range from about 33 d \citep{maas02} up to over 2597 d \citep{winckel99}. The phenomenon of binarity is the most common among RV Tauri variables, and is often reported together with a discussion on possible interaction between components via MT. The idea that the MT leads to the formation of metal-rich RV Tauri appears frequently in the literature \citep{harris84, harris89, winckel99, winckel03, winckel09}. Our simulations further support this idea, but also expose inconsistencies with the canonical evolutionary scenario of RV Tauri stars, which states that RV Tauri stars are progeny of thermally pulsating AGB stars. AGB-core BEPs depart from the AGB before the onset of thermal pulses, and thus their evolutionary status is closer to early post-AGB stars. Moreover, RV Tauri stars are considered supergiants, with radii larger than 100 R$_{\sun}$ \citep{matsuura02}, unlike AGB-core BEPs with much more modest radii distribution ($10-50$ R$_{\sun}$). Altogether, BEP simulation cannot account for the entire sample of binaries with type II Cepheids, but may support the existence of more peculiar ones.

}

\citet{maxted13} reported a stripped red giant core of mass 1.356\,M$_{\sun}$ inside the IS of $\delta$-Scuti variables with a low-mass companion of 0.186\,M$_{\sun}$. We claim that this system represents a different class of pulsators unrelated to BEPs, because the variable star is the more massive component. On the other hand, \citet{maxted14a} reported 17 binary pre-HeWDs, observed during an extremely short period of their evolution while they traverse the HRD horizontally towards higher effective temperatures. Their low masses ($\approx 0.2$ M$_{\sun}$) support the evolutionary scheme, in which post-BEP objects turn into pre-WDs, and encourages further examination of post-BEP evolutionary stages. 

\subsection{The impact of the Kozai mechanism}
Tertiary component in a system of hierarchical configuration can effectively shorten the binary period via Kozai mechanism accompanied by tidal friction, if its orbit inclination exceeds $39\degr$ relative to the orbit of the inner binary \citep*{eggleton06}. As a consequence, the orbital period of inner binary can shrink from more than 30\,d to below 3\,d in roughly 10\,Myr $-$ 1\, Gyr. Theoretical study on the impact of the Kozai mechanism coupled with tidal friction was presented by \citet*{fabrycky07} and the observational evidence of tertiary component in 96\% of field solar-type binaries with orbital periods shorter than 3 d was delivered by \citet{tokovinin06}. We used both sources to provide a crude estimation of the number of BEPs, adjusted for the Kozai mechanism.

The most pronounced increase in the number of close binaries due to the Kozai mechanism should occur for orbital periods in range $0.3 - 8.0$\,d \citep*[Fig. 5 of][]{fabrycky07}. Among our synthetic population only 1\,689 BEP progenitors (6\%) have initial orbital periods within this range. We estimate the number of BEPs to increase by 694 systems within orbital period range $0.3 - 8.0$\,d. Although the relative growth in this period range is large (41\%), this increment has a negligible effect on the entire BEP population, contributing only by 2.5\% to the overall growth of BEP sample.

Hence, we conclude that the inclusion of the Kozai mechanism coupled with tidal fraction has marginal impact on the enlargement of entire BEP population. Although this seems to disagree with P12, who found the Kozai mechanism to noticeably increase the number of BEPs, we recall that their estimations were limited to binaries of initial orbital periods of $2-3$\,d, in which case their entire sample of close binaries got significantly enlarged due to Kozai mechanism; hence BEPs could form from larger sample, and their number increased. In the context of close binaries only, we acknowledge the importance of Kozai mechanism on BEP formation.

\subsection{Contamination of RR Lyrae and Cepheid variables}
BEP variability at first glance can be indistinguishable from genuine RR Lyrae or Cepheid pulsations (P12). Further study \citep{smolec13} shows that subtle characteristics of BEP light curve and radial velocity curve can be used as a tool to separate BEPs from classical pulsators, yet these methods require much attention for individual objects as  light and radial velocity curves must be extracted from abundant observational data and carefully examined. Given scarce observational data or time-saving classification algorithms, BEPs can be assigned by their light curves to RRL or Cepheid variables and thus contaminate the samples of genuine classical pulsators. If the portion of impostors among genuine variables is substantial, they may affect statistics-based calculations involving RRL and Cepheid variables, like age or distance determinations. This issue was already studied by \citet{majaess12} in context of crowded high-surface brightness cores of the clusters. We address this issue in context of binary systems with one component inside the IS.

The contamination value for RRL variables was preliminarily estimated at 0.2\% (P12). We use BEP synthetic population in order to re-evaluate this value for RRLs, and evaluate the contamination value for classical Cepheids.

The contamination value describes the fraction of stars inside the IS, which are in fact RRL or Cepheid impostors:
\begin{equation}
C_\mathrm{var} = 0.5 ~ \frac{N_\mathrm{BEP}}{N_\mathrm{var}} ~ \frac{\Delta t_\mathrm{BEP}}{\Delta t_\mathrm{var}} \times 100\%
\label{eq:contamination}
\end{equation}
where $N_\mathrm{var}$ is the number of RR Lyrae or Cepheid variables, $N_\mathrm{var} = \{ N_\mathrm{RRL},\,N_\mathrm{Cep} \}$, and $N_\mathrm{BEP}$ is the number of BEPs {\color{black}contaminating a given} variable type. 
Average times spent inside the IS by classical variables and BEPs are denoted as $\Delta t_\mathrm{var} = \{ \Delta t_\mathrm{RRL},\,\Delta t_\mathrm{Cep} \}$ and $\Delta t_\mathrm{BEP}$, respectively. 
Because BEPs are always products of binary evolution, the probability of their occurrence needs to be reduced by 50\%, adequately to the number of binary systems among all stars. 

In order to establish the value $N_\mathrm{BEP}$ for Cepheids and RRL variables, we arbitrarily set the threshold luminosity value $\log(L/\mathrm{L_{\sun}}) = 2.0$; BEPs with luminosities below this threshold value are assumed to be RRL impostors, while BEPs above the luminosity threshold mimic Cepheids. According to this threshold, our sample of 29\,386 BEPs is divided into two groups: 25\,091 objects can potentially contaminate Cepheid variables, and 4\,295 objects can potentially contaminate RRL variables. The number of RRL variables, $N_\mathrm{RRL}$, is derived from the synthetic population of 500\,000 single stars with initial masses randomly chosen from the same IMF as BEPs (see Section \ref{sec:calculations}) and under the assumption, that 20\% of stars with initial masses in range $0.8-0.9$ M$_{\sun}$ will become RR~Lyrae stars (P12); outside this range no RRL progenitors are formed. The number of Cepheid variables, $N_\mathrm{Cep}$, is derived analogously, but with different initial mass range ($2.5 - 37.0$ M$_{\sun}$) and fraction of Cepheid progenitors (28\%), which is estimated from a grid of evolutionary tracks constructed with MESA evolutionary code \citep{choi16} for [Fe/H] ranging from $-2.0$ to $0.0$ dex. Above computations are summarized in formulae:
\begin{equation}
\begin{array}{l}
N_\mathrm{RRL} = \xi(0.8, 0.9) \times 0.20\\[0.3em]
N_\mathrm{Cep} = \xi(2.5, 37.0) \times 0.28
\end{array}
\label{eq:Nstar}
\end{equation}
where $\xi$ is the IMF as described in Section \ref{sec:calculations}, values in parenthesis are initial mass ranges, and remaining coefficients inform about the fraction of RRL and Cepheid progenitors within given initial mass range. From the same MESA calculations we extract mean IS crossing time for Cepheids, $\Delta t_\mathrm{Cep} = 0.4$ Myr, weighted by the IMF in a sense that low-mass Cepheid are formed more often and thus their input to the mean IS duration is larger. 
{
\color{black} The average IS crossing time for RRL variables is assumed to be $\Delta t_\mathrm{RRL} = 10$\,Myr \citep[e.g.][]{pritzl02,cassisi04}.
}
The most frequent IS duration for less and more luminous BEPs is 293 kyr and 28 kyr, respectively (see Fig. \ref{fig:is_properties}e). Provided all above information, one can estimate from Eq. \ref{eq:contamination}, that the contamination value for RRL variables is 
{
\color{black} $C_\mathrm{RRL} = 0.8\%$, 
}
and for Cepheid variables is $C_\mathrm{Cep} = 5.0\%$.

Let it be noted here, that contamination values are calculated from crude estimations, and as such are accurate only to an order of magnitude. 
{
\color{black} 
The denominator of the Eq. \ref{eq:contamination} is pivotal for the accuracy of the contamination value, because it holds two parameters, $\Delta t_\mathrm{var}$ and $N_\mathrm{var}$, that span over a wide range of values, and are highly dependent on metallicity, initial mass range, and a fraction of successful pulsators. 
In general, contamination values are not universal and can be re-evaluated for a particular galaxy or cluster with narrower ranges of these parameters. 

While $\Delta t_\mathrm{var}$ and $N_\mathrm{var}$ could be fairly estimated for RRL and classical Cepheid variables, based on the sound evolutionary tracks supported by abundant observational evidences, similar estimations for type II and anomalous Cepheids seem hardly achievable for the time being. Pulsational scenarios thoroughly examined by \citet{bono97d,bono97c} provide IS crossing times for type II and anomalous Cepheids, and their evolutionary tracks in the vicinity of the instability strip are generally agreed upon in the literature, as well as their masses. However, the initial masses are difficult to infer since the evolutionary scenarios are uncertain and can be additionally altered by the presence of a companion \citep[e.g.][]{soszynski08}. Consequently, we lack one piece of information, $N_\mathrm{var}$, to resolve Eq. \ref{eq:contamination} for type II and anomalous Cepheids.
}

The RRL contamination value calculated from above assumptions is 
{
\color{black} four 
}
 times larger that the estimation of P12. This inconsistency results from differences in adopted approaches; the contamination value of P12 was estimated with constraints allowing for only a narrow range of initial masses ($0.9 - 1.4$ M$_{\sun}$) and initial orbital periods ($2-3$ d), while our approach allowed for initial parameters from much wider ranges. Thereby, we not only collected large synthetic sample of BEP objects, but also discovered that BEP physical and orbital parameters are more varied than previously expected. If we adopted the same constraints as P12, our contamination value of RRL stars would converge to the result of P12.

\subsection{Detection limitations}
\label{sec:detection}
The serendipitous discovery of the first BEP was made due to eclipses visible in the RRL-type light curve (P12). Eclipses are not essential for BEP's detection, but they are invaluable pointers to binary systems. Non-eclipsing binaries can be examined spectroscopically with great precision due to radial velocity changes linked to their orbital motion, but this analysis is time-consuming and one cannot afford it in survey approach. Therefore, the most successful way to employ spectroscopic analysis on binary stars, is to choose eclipsing systems. In order to inspect the probability of detecting eclipses in our BEP sample, we constructed for each system a family of light curves with changing inclination value in range $0-90^\circ$, and marked this system as potentially eclipsing if for any inclination value, the drop in the brightness was larger than 5\% of the maximum brightness. 
{
\color{black} As a result, only 485 systems from our sample of 29\,386 (1.65\%) are predisposed to show eclipses due to their favourable physical and orbital properties. Among them, 77 (0.26\%) have luminosities $\log (L / \mathrm{L_{\sun}}) < 2$, so they may be classified as RR Lyrae stars, while the remaining 408 objects (1.39\%) are more luminous, so they lie in the Cepheid domain on the HRD. All these systems show eclipses only if their orbit inclination is at least $77-80^\circ$, with the most common value of minimal inclination of $85^\circ$. This estimation well agrees with the orbit inclination of the first BEP ($83.4^\circ$, P12).
}

One could argue, that binaries without eclipses could still be detectable due to the excess light from the companion, making systems with BEPs considerable brighter compared to the RRL and Cepheid pulsators on the period-luminosity plane. However, the MT could significantly affect BEPs physical structure and pulsational response to the conditions, which trigger pulsations in classical variables. Therefore, the link between the physical properties and the pulsational period of BEPs remains unknown and cannot be claimed to be the same as in classical pulsators, until statistically relevant number of BEPs is discovered.

The next possible way of BEP detection is the ratio of pulsational period change. The change of pulsational period of the first discovered BEP was calculated at $\dot{P} = (-8.4 \pm 2.6) \times 10^{-6}$\,d\,yr$^{-1}$ (P12), which is approximately a hundred times larger than $\dot{P}$ of RRL stars. Based on our results, we confirm that pulsational period of all BEPs of RGB-core and AGB-core types should decrease, as they traverse the HRD horizontally towards higher effective temperatures and smaller radii. We also agree, that the ratio of pulsational period change for BEPs should be considerable larger than for RRL or Cepheid variables, because BEPs cross the IS much faster than classical pulsators. However, we refrain from further approximations of BEP period change until observational data from a larger BEP sample are provided.

\section{Summary and Conclusions}
{
\color{black} We explored a possible occurrence of Binary Evolution Pulsators (BEPs) along the instability strip of RR Lyrae and Cepheid variables, by carrying out synthesis population with StarTrack code on low-mass and metal-rich binary systems. We discovered that BEPs appear among binaries 
}
in a variety of masses, luminosities, radii, and orbital periods. We note that resulted distributions of BEP parameters agree with BEP observed properties, reported by P12. This agreement shows that StarTrack, initially tailored for population synthesis of massive binaries, is a promising tool for an investigation of low-mass interacting binary systems as well.
{
\color{black} Our study provides broad, statistics-based estimates, which can serve as a starting point for more detailed and sophisticated evolutionary modelling.
}

We obtained two contamination values, for RRL and Cepheid stars, which set our exceptions on the fraction of undetected or misclassified BEPs among classical pulsators. These values are estimated at 0.8\% and 5\% for RRL and Cepheid variables, respectively. Although the percentage of undetected BEPs is larger than predicted by P12, statistics-based measurements involving RRL and Cepheid variables should not be affected, as long as they rely on a large sample of variables.

BEPs traverse the IS at fast paces and mostly towards higher effective temperatures. This behavior should be reflected in their large decrease of pulsational periods, which would allow to distinguish them from classical pulsators. Detection of BEPs through eclipses is highly unlikely due to wide orbits and large differences in components' radii. And yet, the first BEP was detected in an eclipsing binary. We suspect that in the future the most BEPs will be exposed by their eclipses, because although eclipses among BEPs are so rare, they unambiguously reveal the nature of a binary.

\section*{Acknowledgements}
We are indebted with Wojtek Dziembowski for valuable discussions. We thank the referee, Giuseppe Bono, whose pertinent comments helped to significantly improve this paper. This work has been carried out within the frame of the Polish National Science Centre grant MAESTRO 2012/06/A/ST9/00269 and the Polish National Science Centre grant PRELUDIUM 2012/07/N/ST9/04246. W.\,G. acknowledges financial support from the BASAL Centro de Astrofisica y Tecnologias Afines (CATA) PFB-06/2007, and from the Millenium Institute of Astrophysics (MAS) of the Iniciativa Cientifica Milenio del Ministerio de Economia, Fomento y Turismo de Chile, project IC120009. This research has made use of NASA's Astrophysics Data System Service.



\bibliographystyle{mnras}
\bibliography{pkarczmarek}

\begin{thebibliography}{}
\makeatletter
\relax
\def\mn@urlcharsother{\let\do\@makeother \do\$\do\&\do\#\do\^\do\_\do\%\do\~}
\def\mn@doi{\begingroup\mn@urlcharsother \@ifnextchar [ {\mn@doi@}
  {\mn@doi@[]}}
\def\mn@doi@[#1]#2{\def\@tempa{#1}\ifx\@tempa\@empty \href
  {http://dx.doi.org/#2} {doi:#2}\else \href {http://dx.doi.org/#2} {#1}\fi
  \endgroup}
\def\mn@eprint#1#2{\mn@eprint@#1:#2::\@nil}
\def\mn@eprint@arXiv#1{\href {http://arxiv.org/abs/#1} {{\tt arXiv:#1}}}
\def\mn@eprint@dblp#1{\href {http://dblp.uni-trier.de/rec/bibtex/#1.xml}
  {dblp:#1}}
\def\mn@eprint@#1:#2:#3:#4\@nil{\def\@tempa {#1}\def\@tempb {#2}\def\@tempc
  {#3}\ifx \@tempc \@empty \let \@tempc \@tempb \let \@tempb \@tempa \fi \ifx
  \@tempb \@empty \def\@tempb {arXiv}\fi \@ifundefined
  {mn@eprint@\@tempb}{\@tempb:\@tempc}{\expandafter \expandafter \csname
  mn@eprint@\@tempb\endcsname \expandafter{\@tempc}}}

\bibitem[\protect\citeauthoryear{{Abate}, {Pols}, {Izzard}, {Mohamed}  \& {de
  Mink}}{{Abate} et~al.}{2013}]{abate13}
{Abate} C.,  {Pols} O.~R.,  {Izzard} R.~G.,  {Mohamed} S.~S.,   {de Mink}
  S.~E.,  2013, \mn@doi [\aap] {10.1051/0004-6361/201220007}, \href
  {http://adsabs.harvard.edu/abs/2013A%26A...552A..26A} {552, A26}

\bibitem[\protect\citeauthoryear{{Abt}}{{Abt}}{1983}]{abt83}
{Abt} H.~A.,  1983, \mn@doi [\araa] {10.1146/annurev.aa.21.090183.002015},
  \href {http://adsabs.harvard.edu/abs/1983ARA%26A..21..343A} {21, 343}

\bibitem[\protect\citeauthoryear{{Belczynski}, {Kalogera}  \&
  {Bulik}}{{Belczynski} et~al.}{2002}]{belczynski02}
{Belczynski} K.,  {Kalogera} V.,   {Bulik} T.,  2002, \mn@doi [\apj]
  {10.1086/340304}, \href {http://adsabs.harvard.edu/abs/2002ApJ...572..407B}
  {572, 407}

\bibitem[\protect\citeauthoryear{{Belczynski}, {Kalogera}, {Rasio}, {Taam},
  {Zezas}, {Bulik}, {Maccarone}  \& {Ivanova}}{{Belczynski}
  et~al.}{2008}]{belczynski08}
{Belczynski} K.,  {Kalogera} V.,  {Rasio} F.~A.,  {Taam} R.~E.,  {Zezas} A.,
  {Bulik} T.,  {Maccarone} T.~J.,   {Ivanova} N.,  2008, \mn@doi [\apjs]
  {10.1086/521026}, \href {http://adsabs.harvard.edu/abs/2008ApJS..174..223B}
  {174, 223}

\bibitem[\protect\citeauthoryear{{Bono}, {Caputo}, {Santolamazza}, {Cassisi}
  \& {Piersimoni}}{{Bono} et~al.}{1997a}]{bono97d}
{Bono} G.,  {Caputo} F.,  {Santolamazza} P.,  {Cassisi} S.,   {Piersimoni} A.,
  1997a, \mn@doi [\aj] {10.1086/118431}, \href
  {http://adsabs.harvard.edu/abs/1997AJ....113.2209B} {113, 2209}

\bibitem[\protect\citeauthoryear{{Bono}, {Caputo}  \& {Santolamazza}}{{Bono}
  et~al.}{1997b}]{bono97c}
{Bono} G.,  {Caputo} F.,   {Santolamazza} P.,  1997b, \aap, \href
  {http://adsabs.harvard.edu/abs/1997A%26A...317..171B} {317, 171}

\bibitem[\protect\citeauthoryear{{Bono}, {Caputo}, {Cassisi}, {Castellani}  \&
  {Marconi}}{{Bono} et~al.}{1997c}]{bono97}
{Bono} G.,  {Caputo} F.,  {Cassisi} S.,  {Castellani} V.,   {Marconi} M.,
  1997c, \apj, \href {http://adsabs.harvard.edu/abs/1997ApJ...479..279B} {479,
  279}

\bibitem[\protect\citeauthoryear{{Bono}, {Caputo}, {Cassisi}, {Incerpi}  \&
  {Marconi}}{{Bono} et~al.}{1997d}]{bono97a}
{Bono} G.,  {Caputo} F.,  {Cassisi} S.,  {Incerpi} R.,   {Marconi} M.,  1997d,
  \apj, \href {http://adsabs.harvard.edu/abs/1997ApJ...483..811B} {483, 811}

\bibitem[\protect\citeauthoryear{{Bono}, {Caputo}, {Castellani}, {Marconi},
  {Storm}  \& {Degl'Innocenti}}{{Bono} et~al.}{2003}]{bono03}
{Bono} G.,  {Caputo} F.,  {Castellani} V.,  {Marconi} M.,  {Storm} J.,
  {Degl'Innocenti} S.,  2003, \mn@doi [\mnras]
  {10.1046/j.1365-8711.2003.06878.x}, \href
  {http://adsabs.harvard.edu/abs/2003MNRAS.344.1097B} {344, 1097}

\bibitem[\protect\citeauthoryear{{Brown}, {Kilic}, {Brown}  \&
  {Kenyon}}{{Brown} et~al.}{2011}]{brown11}
{Brown} J.~M.,  {Kilic} M.,  {Brown} W.~R.,   {Kenyon} S.~J.,  2011, \mn@doi
  [\apj] {10.1088/0004-637X/730/2/67}, \href
  {http://adsabs.harvard.edu/abs/2011ApJ...730...67B} {730, 67}

\bibitem[\protect\citeauthoryear{{Cassisi}, {Castellani}, {Caputo}  \&
  {Castellani}}{{Cassisi} et~al.}{2004}]{cassisi04}
{Cassisi} S.,  {Castellani} M.,  {Caputo} F.,   {Castellani} V.,  2004, \mn@doi
  [\aap] {10.1051/0004-6361:20041048}, \href
  {http://adsabs.harvard.edu/abs/2004A%26A...426..641C} {426, 641}

\bibitem[\protect\citeauthoryear{{Choi}, {Dotter}, {Conroy}, {Cantiello},
  {Paxton}  \& {Johnson}}{{Choi} et~al.}{2016}]{choi16}
{Choi} J.,  {Dotter} A.,  {Conroy} C.,  {Cantiello} M.,  {Paxton} B.,
  {Johnson} B.~D.,  2016, \mn@doi [\apj] {10.3847/0004-637X/823/2/102}, \href
  {http://adsabs.harvard.edu/abs/2016ApJ...823..102C} {823, 102}

\bibitem[\protect\citeauthoryear{{Christy}}{{Christy}}{1966}]{christy66}
{Christy} R.~F.,  1966, \mn@doi [\apj] {10.1086/148593}, \href
  {http://adsabs.harvard.edu/abs/1966ApJ...144..108C} {144, 108}

\bibitem[\protect\citeauthoryear{{D'Cruz}, {Dorman}, {Rood}  \&
  {O'Connell}}{{D'Cruz} et~al.}{1996}]{dcruz96}
{D'Cruz} N.~L.,  {Dorman} B.,  {Rood} R.~T.,   {O'Connell} R.~W.,  1996,
  \mn@doi [\apj] {10.1086/177515}, \href
  {http://adsabs.harvard.edu/abs/1996ApJ...466..359D} {466, 359}

\bibitem[\protect\citeauthoryear{{Driebe}, {Schoenberner}, {Bloecker}  \&
  {Herwig}}{{Driebe} et~al.}{1998}]{driebe98}
{Driebe} T.,  {Schoenberner} D.,  {Bloecker} T.,   {Herwig} F.,  1998, \aap,
  \href {http://adsabs.harvard.edu/abs/1998A%26A...339..123D} {339, 123}

\bibitem[\protect\citeauthoryear{{Eggleton} \& {Kisseleva-Eggleton}}{{Eggleton}
  \& {Kisseleva-Eggleton}}{2006}]{eggleton06}
{Eggleton} P.~P.,  {Kisseleva-Eggleton} L.,  2006, \mn@doi [\apss]
  {10.1007/s10509-006-9078-z}, \href
  {http://adsabs.harvard.edu/abs/2006Ap%26SS.304...75E} {304, 75}

\bibitem[\protect\citeauthoryear{{Fabrycky} \& {Tremaine}}{{Fabrycky} \&
  {Tremaine}}{2007}]{fabrycky07}
{Fabrycky} D.,  {Tremaine} S.,  2007, \mn@doi [\apj] {10.1086/521702}, \href
  {http://adsabs.harvard.edu/abs/2007ApJ...669.1298F} {669, 1298}

\bibitem[\protect\citeauthoryear{{Gaulme}, {Jackiewicz}, {Appourchaux}  \&
  {Mosser}}{{Gaulme} et~al.}{2014}]{gaulme14}
{Gaulme} P.,  {Jackiewicz} J.,  {Appourchaux} T.,   {Mosser} B.,  2014, \mn@doi
  [\apj] {10.1088/0004-637X/785/1/5}, \href
  {http://adsabs.harvard.edu/abs/2014ApJ...785....5G} {785, 5}

\bibitem[\protect\citeauthoryear{{Gieren} et~al.,}{{Gieren}
  et~al.}{2014}]{gieren14}
{Gieren} W.,  et~al., 2014, \mn@doi [\apj] {10.1088/0004-637X/786/2/80}, \href
  {http://adsabs.harvard.edu/abs/2014ApJ...786...80G} {786, 80}

\bibitem[\protect\citeauthoryear{{Guzik}, {Bradley}, {Jackiewicz},
  {Uytterhoeven}  \& {Kinemuchi}}{{Guzik} et~al.}{2014}]{guzik14}
{Guzik} J.~A.,  {Bradley} P.~A.,  {Jackiewicz} J.,  {Uytterhoeven} K.,
  {Kinemuchi} K.,  2014, in {Guzik} J.~A.,  {Chaplin} W.~J.,  {Handler} G.,
  {Pigulski} A.,  eds,  IAU Symposium Vol. 301, IAU Symposium. pp 63--66,
  \mn@doi{10.1017/S1743921313014099}

\bibitem[\protect\citeauthoryear{{Hajdu}, {Catelan}, {Jurcsik},
  {D{\'e}k{\'a}ny}, {Drake}  \& {Marquette}}{{Hajdu} et~al.}{2015}]{hajdu15}
{Hajdu} G.,  {Catelan} M.,  {Jurcsik} J.,  {D{\'e}k{\'a}ny} I.,  {Drake} A.~J.,
    {Marquette} J.-B.,  2015, \mn@doi [\mnras] {10.1093/mnrasl/slv024}, \href
  {http://adsabs.harvard.edu/abs/2015MNRAS.449L.113H} {449, L113}

\bibitem[\protect\citeauthoryear{{Han}, {Tout}  \& {Eggleton}}{{Han}
  et~al.}{2000}]{han00}
{Han} Z.,  {Tout} C.~A.,   {Eggleton} P.~P.,  2000, \mn@doi [\mnras]
  {10.1046/j.1365-8711.2000.03839.x}, \href
  {http://adsabs.harvard.edu/abs/2000MNRAS.319..215H} {319, 215}

\bibitem[\protect\citeauthoryear{{Harris} \& {Welch}}{{Harris} \&
  {Welch}}{1989}]{harris89}
{Harris} H.~C.,  {Welch} D.~L.,  1989, \mn@doi [\aj] {10.1086/115190}, \href
  {http://adsabs.harvard.edu/abs/1989AJ.....98..981H} {98, 981}

\bibitem[\protect\citeauthoryear{{Harris}, {Olszewski}  \&
  {Wallerstein}}{{Harris} et~al.}{1984}]{harris84}
{Harris} H.~C.,  {Olszewski} E.~W.,   {Wallerstein} G.,  1984, \mn@doi [\aj]
  {10.1086/113489}, \href {http://adsabs.harvard.edu/abs/1984AJ.....89..119H}
  {89, 119}

\bibitem[\protect\citeauthoryear{{Heggie}}{{Heggie}}{1975}]{heggie75}
{Heggie} D.~C.,  1975, \mnras, \href
  {http://adsabs.harvard.edu/abs/1975MNRAS.173..729H} {173, 729}

\bibitem[\protect\citeauthoryear{{Hermes} et~al.,}{{Hermes}
  et~al.}{2013}]{hermes13}
{Hermes} J.~J.,  et~al., 2013, \mn@doi [\mnras] {10.1093/mnras/stt1835}, \href
  {http://adsabs.harvard.edu/abs/2013MNRAS.436.3573H} {436, 3573}

\bibitem[\protect\citeauthoryear{{Hurley}, {Pols}  \& {Tout}}{{Hurley}
  et~al.}{2000}]{hurley00}
{Hurley} J.~R.,  {Pols} O.~R.,   {Tout} C.~A.,  2000, \mn@doi [\mnras]
  {10.1046/j.1365-8711.2000.03426.x}, \href
  {http://adsabs.harvard.edu/abs/2000MNRAS.315..543H} {315, 543}

\bibitem[\protect\citeauthoryear{{Hurley}, {Tout}  \& {Pols}}{{Hurley}
  et~al.}{2002}]{hurley02}
{Hurley} J.~R.,  {Tout} C.~A.,   {Pols} O.~R.,  2002, \mn@doi [\mnras]
  {10.1046/j.1365-8711.2002.05038.x}, \href
  {http://adsabs.harvard.edu/abs/2002MNRAS.329..897H} {329, 897}

\bibitem[\protect\citeauthoryear{{Iben} \& {Livio}}{{Iben} \&
  {Livio}}{1993}]{iben93}
{Iben} Jr. I.,  {Livio} M.,  1993, \mn@doi [\pasp] {10.1086/133321}, \href
  {http://adsabs.harvard.edu/abs/1993PASP..105.1373I} {105, 1373}

\bibitem[\protect\citeauthoryear{{J{\o}rgensen} \& {Petersen}}{{J{\o}rgensen}
  \& {Petersen}}{1967}]{jorgensen67}
{J{\o}rgensen} H.~E.,  {Petersen} J.~O.,  1967, \zap, \href
  {http://adsabs.harvard.edu/abs/1967ZA.....67..377J} {67, 377}

\bibitem[\protect\citeauthoryear{{Karczmarek}}{{Karczmarek}}{2012}]{karczmarek12}
{Karczmarek} P.,  2012, Advances in Astronomy and Space Physics, \href
  {http://adsabs.harvard.edu/abs/2012AASP....2..135K} {2, 135}

\bibitem[\protect\citeauthoryear{{Kobulnicky} \& {Fryer}}{{Kobulnicky} \&
  {Fryer}}{2007}]{kobulnicky07}
{Kobulnicky} H.~A.,  {Fryer} C.~L.,  2007, \mn@doi [\apj] {10.1086/522073},
  \href {http://adsabs.harvard.edu/abs/2007ApJ...670..747K} {670, 747}

\bibitem[\protect\citeauthoryear{{Kroupa} \& {Weidner}}{{Kroupa} \&
  {Weidner}}{2003}]{kroupa03}
{Kroupa} P.,  {Weidner} C.,  2003, \mn@doi [\apj] {10.1086/379105}, \href
  {http://adsabs.harvard.edu/abs/2003ApJ...598.1076K} {598, 1076}

\bibitem[\protect\citeauthoryear{{Liakos}, {Niarchos}, {Soydugan}  \&
  {Zasche}}{{Liakos} et~al.}{2012}]{liakos12}
{Liakos} A.,  {Niarchos} P.,  {Soydugan} E.,   {Zasche} P.,  2012, \mn@doi
  [\mnras] {10.1111/j.1365-2966.2012.20704.x}, \href
  {http://adsabs.harvard.edu/abs/2012MNRAS.422.1250L} {422, 1250}

\bibitem[\protect\citeauthoryear{{Lynas-Gray}}{{Lynas-Gray}}{2013}]{lynas13}
{Lynas-Gray} A.~E.,  2013, in {Shibahashi} H.,  {Lynas-Gray} A.~E.,  eds,
  Astronomical Society of the Pacific Conference Series Vol. 479, Progress in
  Physics of the Sun and Stars: A New Era in Helio- and Asteroseismology.
  p.~273

\bibitem[\protect\citeauthoryear{{Maas}, {Van Winckel}  \& {Waelkens}}{{Maas}
  et~al.}{2002}]{maas02}
{Maas} T.,  {Van Winckel} H.,   {Waelkens} C.,  2002, \mn@doi [\aap]
  {10.1051/0004-6361:20020209}, \href
  {http://adsabs.harvard.edu/abs/2002A%26A...386..504M} {386, 504}

\bibitem[\protect\citeauthoryear{{Majaess}, {Turner}, {Gieren}  \&
  {Lane}}{{Majaess} et~al.}{2012}]{majaess12}
{Majaess} D.,  {Turner} D.,  {Gieren} W.,   {Lane} D.,  2012, \mn@doi [\apjl]
  {10.1088/2041-8205/752/1/L10}, \href
  {http://adsabs.harvard.edu/abs/2012ApJ...752L..10M} {752, L10}

\bibitem[\protect\citeauthoryear{{Mathieu}}{{Mathieu}}{1994}]{mathieu94}
{Mathieu} R.~D.,  1994, \mn@doi [\araa] {10.1146/annurev.aa.32.090194.002341},
  \href {http://adsabs.harvard.edu/abs/1994ARA%26A..32..465M} {32, 465}

\bibitem[\protect\citeauthoryear{{Matsuura}, {Yamamura}, {Zijlstra}  \&
  {Bedding}}{{Matsuura} et~al.}{2002}]{matsuura02}
{Matsuura} M.,  {Yamamura} I.,  {Zijlstra} A.~A.,   {Bedding} T.~R.,  2002,
  \mn@doi [\aap] {10.1051/0004-6361:20020391}, \href
  {http://adsabs.harvard.edu/abs/2002A%26A...387.1022M} {387, 1022}

\bibitem[\protect\citeauthoryear{{Maxted} et~al.,}{{Maxted}
  et~al.}{2011}]{maxted11}
{Maxted} P.~F.~L.,  et~al., 2011, \mn@doi [\mnras]
  {10.1111/j.1365-2966.2011.19567.x}, \href
  {http://adsabs.harvard.edu/abs/2011MNRAS.418.1156M} {418, 1156}

\bibitem[\protect\citeauthoryear{{Maxted} et~al.,}{{Maxted}
  et~al.}{2013}]{maxted13}
{Maxted} P.~F.~L.,  et~al., 2013, \mn@doi [\nat] {10.1038/nature12192}, \href
  {http://adsabs.harvard.edu/abs/2013Natur.498..463M} {498, 463}

\bibitem[\protect\citeauthoryear{{Maxted} et~al.,}{{Maxted}
  et~al.}{2014}]{maxted14a}
{Maxted} P.~F.~L.,  et~al., 2014, \mn@doi [\mnras] {10.1093/mnras/stt2007},
  \href {http://adsabs.harvard.edu/abs/2014MNRAS.437.1681M} {437, 1681}

\bibitem[\protect\citeauthoryear{{Mohamed} \& {Podsiadlowski}}{{Mohamed} \&
  {Podsiadlowski}}{2012}]{mohamed12}
{Mohamed} S.,  {Podsiadlowski} P.,  2012, Baltic Astronomy, \href
  {http://adsabs.harvard.edu/abs/2012BaltA..21...88M} {21, 88}

\bibitem[\protect\citeauthoryear{{Murphy}, {Bedding}, {Niemczura}, {Kurtz}  \&
  {Smalley}}{{Murphy} et~al.}{2015}]{murphy15}
{Murphy} S.~J.,  {Bedding} T.~R.,  {Niemczura} E.,  {Kurtz} D.~W.,   {Smalley}
  B.,  2015, \mn@doi [\mnras] {10.1093/mnras/stu2749}, \href
  {http://adsabs.harvard.edu/abs/2015MNRAS.447.3948M} {447, 3948}

\bibitem[\protect\citeauthoryear{{Neilson}, {Schneider}, {Izzard}, {Evans}  \&
  {Langer}}{{Neilson} et~al.}{2015}]{neilson15}
{Neilson} H.~R.,  {Schneider} F.~R.~N.,  {Izzard} R.~G.,  {Evans} N.~R.,
  {Langer} N.,  2015, \mn@doi [\aap] {10.1051/0004-6361/201424408}, \href
  {http://adsabs.harvard.edu/abs/2015A%26A...574A...2N} {574, A2}

\bibitem[\protect\citeauthoryear{{O'Connell}}{{O'Connell}}{1999}]{oconnell99}
{O'Connell} R.~W.,  1999, \mn@doi [\araa] {10.1146/annurev.astro.37.1.603},
  \href {http://adsabs.harvard.edu/abs/1999ARA%26A..37..603O} {37, 603}

\bibitem[\protect\citeauthoryear{{Pavlovskii} \& {Ivanova}}{{Pavlovskii} \&
  {Ivanova}}{2015}]{pavlovskii15}
{Pavlovskii} K.,  {Ivanova} N.,  2015, \mn@doi [\mnras] {10.1093/mnras/stv619},
  \href {http://adsabs.harvard.edu/abs/2015MNRAS.449.4415P} {449, 4415}

\bibitem[\protect\citeauthoryear{{Paxton} et~al.,}{{Paxton}
  et~al.}{2015}]{paxton15}
{Paxton} B.,  et~al., 2015, \mn@doi [\apjs] {10.1088/0067-0049/220/1/15}, \href
  {http://adsabs.harvard.edu/abs/2015ApJS..220...15P} {220, 15}

\bibitem[\protect\citeauthoryear{{Pietrukowicz} et~al.,}{{Pietrukowicz}
  et~al.}{2015}]{pietrukowicz15}
{Pietrukowicz} P.,  et~al., 2015, \mn@doi [\apj] {10.1088/0004-637X/811/2/113},
  \href {http://adsabs.harvard.edu/abs/2015ApJ...811..113P} {811, 113}

\bibitem[\protect\citeauthoryear{{Pietrzy{\'n}ski} et~al.,}{{Pietrzy{\'n}ski}
  et~al.}{2010}]{pietrzynski10}
{Pietrzy{\'n}ski} G.,  et~al., 2010, \mn@doi [\nat] {10.1038/nature09598},
  \href {http://adsabs.harvard.edu/abs/2010Natur.468..542P} {468, 542}

\bibitem[\protect\citeauthoryear{{Pietrzy{\'n}ski} et~al.,}{{Pietrzy{\'n}ski}
  et~al.}{2011}]{pietrzynski11}
{Pietrzy{\'n}ski} G.,  et~al., 2011, \mn@doi [\apjl]
  {10.1088/2041-8205/742/2/L20}, \href
  {http://adsabs.harvard.edu/abs/2011ApJ...742L..20P} {742, L20}

\bibitem[\protect\citeauthoryear{Pietrzy{\'n}ski et~al.,}{Pietrzy{\'n}ski
  et~al.}{2012}]{pietrzynski12}
Pietrzy{\'n}ski G.,  et~al., 2012, \mn@doi [Nature] {10.1038/nature10966}, 484,
  75 (P12)

\bibitem[\protect\citeauthoryear{{Pilecki} et~al.,}{{Pilecki}
  et~al.}{2013}]{pilecki13}
{Pilecki} B.,  et~al., 2013, \mn@doi [\mnras] {10.1093/mnras/stt1529}, \href
  {http://adsabs.harvard.edu/abs/2013MNRAS.436..953P} {436, 953}

\bibitem[\protect\citeauthoryear{{Pilecki} et~al.,}{{Pilecki}
  et~al.}{2015}]{pilecki15}
{Pilecki} B.,  et~al., 2015, \mn@doi [\apj] {10.1088/0004-637X/806/1/29}, \href
  {http://adsabs.harvard.edu/abs/2015ApJ...806...29P} {806, 29}

\bibitem[\protect\citeauthoryear{{Podsiadlowski}, {Rappaport}  \&
  {Pfahl}}{{Podsiadlowski} et~al.}{2002}]{podsiadlowski02}
{Podsiadlowski} P.,  {Rappaport} S.,   {Pfahl} E.~D.,  2002, \mn@doi [\apj]
  {10.1086/324686}, \href {http://adsabs.harvard.edu/abs/2002ApJ...565.1107P}
  {565, 1107}

\bibitem[\protect\citeauthoryear{{Pols}, {Schr{\"o}der}, {Hurley}, {Tout}  \&
  {Eggleton}}{{Pols} et~al.}{1998}]{pols98}
{Pols} O.~R.,  {Schr{\"o}der} K.-P.,  {Hurley} J.~R.,  {Tout} C.~A.,
  {Eggleton} P.~P.,  1998, \mn@doi [\mnras] {10.1046/j.1365-8711.1998.01658.x},
  \href {http://adsabs.harvard.edu/abs/1998MNRAS.298..525P} {298, 525}

\bibitem[\protect\citeauthoryear{{Prada Moroni} \& {Straniero}}{{Prada Moroni}
  \& {Straniero}}{2009}]{pradamoroni09}
{Prada Moroni} P.~G.,  {Straniero} O.,  2009, \mn@doi [\aap]
  {10.1051/0004-6361/200912847}, \href
  {http://adsabs.harvard.edu/abs/2009A%26A...507.1575P} {507, 1575}

\bibitem[\protect\citeauthoryear{{Pritzl}, {Smith}, {Catelan}  \&
  {Sweigart}}{{Pritzl} et~al.}{2002}]{pritzl02}
{Pritzl} B.~J.,  {Smith} H.~A.,  {Catelan} M.,   {Sweigart} A.~V.,  2002,
  \mn@doi [\aj] {10.1086/341381}, \href
  {http://adsabs.harvard.edu/abs/2002AJ....124..949P} {124, 949}

\bibitem[\protect\citeauthoryear{{Ritter}}{{Ritter}}{1879}]{ritter79}
{Ritter} A.,  1879, Wiedemanns Annalen, VIII, 173

\bibitem[\protect\citeauthoryear{{Ritter}}{{Ritter}}{1988}]{ritter88}
{Ritter} H.,  1988, \aap, \href
  {http://adsabs.harvard.edu/abs/1988A%26A...202...93R} {202, 93}

\bibitem[\protect\citeauthoryear{{Saio} \& {Gautschy}}{{Saio} \&
  {Gautschy}}{1998}]{saio98}
{Saio} H.,  {Gautschy} A.,  1998, \mn@doi [\apj] {10.1086/305544}, \href
  {http://adsabs.harvard.edu/abs/1998ApJ...498..360S} {498, 360}

\bibitem[\protect\citeauthoryear{{Smolec} et~al.,}{{Smolec}
  et~al.}{2013}]{smolec13}
{Smolec} R.,  et~al., 2013, \mn@doi [\mnras] {10.1093/mnras/sts258}, \href
  {http://adsabs.harvard.edu/abs/2013MNRAS.428.3034S} {428, 3034}

\bibitem[\protect\citeauthoryear{{Soszy{\'n}ski} et~al.,}{{Soszy{\'n}ski}
  et~al.}{2008}]{soszynski08}
{Soszy{\'n}ski} I.,  et~al., 2008, \actaa, \href
  {http://adsabs.harvard.edu/abs/2008AcA....58..293S} {58, 293}

\bibitem[\protect\citeauthoryear{{Soszy{\'n}ski}, {Udalski}, {Szyma{\'n}ski},
  {Kubiak}, {Pietrzy{\'n}ski}, {Wyrzykowski}, {Ulaczyk}  \&
  {Poleski}}{{Soszy{\'n}ski} et~al.}{2010}]{soszynski10}
{Soszy{\'n}ski} I.,  {Udalski} A.,  {Szyma{\'n}ski} M.~K.,  {Kubiak} M.,
  {Pietrzy{\'n}ski} G.,  {Wyrzykowski} {\L}.,  {Ulaczyk} K.,   {Poleski} R.,
  2010, \actaa, \href {http://adsabs.harvard.edu/abs/2010AcA....60...91S} {60,
  91}

\bibitem[\protect\citeauthoryear{{Soszy{\'n}ski} et~al.,}{{Soszy{\'n}ski}
  et~al.}{2011}]{soszynski11}
{Soszy{\'n}ski} I.,  et~al., 2011, \actaa, \href
  {http://adsabs.harvard.edu/abs/2011AcA....61..285S} {61, 285}

\bibitem[\protect\citeauthoryear{{Soydugan}, {Soydugan}, {{\c S}eny{\"u}z},
  {T{\"u}ys{\"u}z}, {Baki{\c s}}, {Bilir}, {{\c C}i{\c c}ek}  \&
  {Demircan}}{{Soydugan} et~al.}{2010}]{soydugan10}
{Soydugan} E.,  {Soydugan} F.,  {{\c S}eny{\"u}z} T.,  {T{\"u}ys{\"u}z} M.,
  {Baki{\c s}} V.,  {Bilir} S.,  {{\c C}i{\c c}ek} C.,   {Demircan} O.,  2010,
  in {Pr{\v s}a} A.,  {Zejda} M.,  eds,  Astronomical Society of the Pacific
  Conference Series Vol. 435, Binaries - Key to Comprehension of the Universe.
  p.~331

\bibitem[\protect\citeauthoryear{{St{\k e}pie{\'n}} \& {Gazeas}}{{St{\k
  e}pie{\'n}} \& {Gazeas}}{2012}]{stepien12}
{St{\k e}pie{\'n}} K.,  {Gazeas} K.,  2012, \actaa, \href
  {http://adsabs.harvard.edu/abs/2012AcA....62..153S} {62, 153}

\bibitem[\protect\citeauthoryear{{Szabados} \& {Pont}}{{Szabados} \&
  {Pont}}{1998}]{szabados98}
{Szabados} L.,  {Pont} F.,  1998, \mn@doi [\aaps] {10.1051/aas:1998103}, \href
  {http://adsabs.harvard.edu/abs/1998A%26AS..133...51S} {133, 51}

\bibitem[\protect\citeauthoryear{{Tokovinin}, {Thomas}, {Sterzik}  \&
  {Udry}}{{Tokovinin} et~al.}{2006}]{tokovinin06}
{Tokovinin} A.,  {Thomas} S.,  {Sterzik} M.,   {Udry} S.,  2006, \mn@doi [\aap]
  {10.1051/0004-6361:20054427}, \href
  {http://adsabs.harvard.edu/abs/2006A%26A...450..681T} {450, 681}

\bibitem[\protect\citeauthoryear{{Van Winckel}}{{Van
  Winckel}}{2003}]{winckel03}
{Van Winckel} H.,  2003, \mn@doi [\araa]
  {10.1146/annurev.astro.41.071601.170018}, \href
  {http://adsabs.harvard.edu/abs/2003ARA%26A..41..391V} {41, 391}

\bibitem[\protect\citeauthoryear{{Van Winckel}, {Waelkens}, {Fernie}  \&
  {Waters}}{{Van Winckel} et~al.}{1999}]{winckel99}
{Van Winckel} H.,  {Waelkens} C.,  {Fernie} J.~D.,   {Waters} L.~B.~F.~M.,
  1999, \aap, \href {http://adsabs.harvard.edu/abs/1999A%26A...343..202V} {343,
  202}

\bibitem[\protect\citeauthoryear{{Walker} \& {Terndrup}}{{Walker} \&
  {Terndrup}}{1991}]{walker91}
{Walker} A.~R.,  {Terndrup} D.~M.,  1991, \mn@doi [\apj] {10.1086/170411},
  \href {http://adsabs.harvard.edu/abs/1991ApJ...378..119W} {378, 119}

\bibitem[\protect\citeauthoryear{{Wallerstein}, {Gomez}  \&
  {Huang}}{{Wallerstein} et~al.}{2012}]{wallerstein12}
{Wallerstein} G.,  {Gomez} T.,   {Huang} W.,  2012, \mn@doi [\apss]
  {10.1007/s10509-012-1033-6}, \href
  {http://adsabs.harvard.edu/abs/2012Ap%26SS.341...89W} {341, 89}

\bibitem[\protect\citeauthoryear{Yakut \& Eggleton}{Yakut \&
  Eggleton}{2005}]{yakut05}
Yakut K.,  Eggleton P.~P.,  2005, The Astrophysical Journal, 629, 1055

\bibitem[\protect\citeauthoryear{{van Winckel} et~al.,}{{van Winckel}
  et~al.}{2009}]{winckel09}
{van Winckel} H.,  et~al., 2009, \mn@doi [\aap] {10.1051/0004-6361/200912332},
  \href {http://adsabs.harvard.edu/abs/2009A%26A...505.1221V} {505, 1221}

\makeatother
\end{thebibliography}




\bsp	
\label{lastpage}
\end{document}